\documentclass[conference]{IEEEtran}

\IEEEoverridecommandlockouts
\usepackage{cleveref}
\usepackage[english]{babel}
\usepackage{balance}
\renewcommand{\baselinestretch}{.942}
\usepackage[utf8]{inputenc} 
\usepackage[labelfont=bf]{caption}
\usepackage[font=small,skip=3pt]{caption}
\usepackage[nodisplayskipstretch]{setspace}
\usepackage{booktabs}
\usepackage[table,xcdraw]{xcolor}
\usepackage{pgfplots}
\usepackage{cite}
\usepackage{multirow}
\usepackage{amsmath,amssymb,amsfonts}
\usepackage{algorithmic}
\usepackage{subfig}
\usepackage{graphicx}
\usepackage{caption}
\usepackage{textcomp}
\usepackage{xcolor}
\usepackage{url}
\usepackage{tablefootnote}
\usepackage{threeparttable}
\DeclareUnicodeCharacter{00D5}{\~y}
\def\BibTeX{{\rm B\kern-.05em{\sc i\kern-.025em b}\kern-.08em
    T\kern-.1667em\lower.7ex\hbox{E}\kern-.125emX}}
\pgfplotsset{compat=1.14}
\begin{document}
\setlength{\belowdisplayskip}{6pt} \setlength{\belowdisplayshortskip}{6pt}
\setlength{\abovedisplayskip}{6pt} \setlength{\abovedisplayshortskip}{6pt}
\setlength{\floatsep}{6pt}
\setlength{\textfloatsep}{6pt}
\setlength{\dbltextfloatsep}{6pt}
\setlength{\abovecaptionskip}{6pt}
\title{ Attack of the Genes: Finding Keys and Parameters of Locked Analog ICs Using Genetic Algorithm} 
{\footnotesize }

\author{\IEEEauthorblockN{Rabin Yu Acharya, Sreeja Chowdhury, Fatemeh Ganji, and Domenic Forte}

\IEEEauthorblockA{Department of Electrical and Computer Engineering, University of Florida, Gainesville, Florida, USA}
\IEEEauthorblockA{\{rabin.acharya, sreejachowdhury, fganji\}@ufl.edu, dforte@ece.ufl.edu}
}
\maketitle

\begin{abstract}
Hardware intellectual property (IP) theft is a major issue in today's globalized supply chain. To address it, numerous logic locking and obfuscation techniques have been proposed. While locking initially focused on digital integrated circuits (ICs), there have been recent attempts to extend it to analog ICs, which are easier to reverse engineer and to copy than digital ICs.  In this paper, we use algorithms based on evolutionary strategies to investigate the security of analog obfuscation/locking techniques. We present a genetic algorithm (GA) approach which is capable of completely breaking a locked analog circuit by finding either its obfuscation key or its obfuscated parameters.  We implement both the GA attack as well as a more na\"ive satisfiability modulo theory (SMT)-based attack on common analog benchmark circuits obfuscated by combinational locking and parameter biasing. We find that GA attack can unlock all the circuits using only the locked netlist and an unlocked chip in minutes. On the other hand, while the SMT attack converges faster, it requires circuit specification to execute and it also returns multiple keys that need to be brute-forced by a post-processing step. We also discuss how the GA attack can generalize to other recent analog locking techniques not tested in the paper\footnote{\textsuperscript{\textcopyright}2020 IEEE. Personal use of this material is permitted. Permission from IEEE must be obtained for all other uses, in any current or future media, including reprinting/republishing this material for advertising or promotional purposes, creating new collective works, for resale or redistribution to servers or lists, or reuse of any copyrighted component of this work in other works}.
\end{abstract}
\section{Introduction}
With the increasing sophistication of integrated circuits (ICs), the electronic design and fabrication industry has shifted from a vertical business model to a horizontal one, enabling cheaper manufacturing and lesser time to market. The new model involves third-party manufacturers and vendors, thus creating security risks such as counterfeiting~\cite{book:counterfeit_IC}, hardware Trojans~\cite{tehranipoor2010survey}, etc. Compared to digital ICs, analog and mixed-signal (AMS) ICs are considered an easier target for reverse engineering, piracy, and counterfeiting~\cite{YM_analog_survey,Alam2018} due to their large technology node size, low transistor counts, and lesser number input/output (I/O) pins. According to IHS, AMS ICs form the largest category of counterfeit ICs and contribute to 25\% of the total number of reported instances~\cite{IHS_data}. 
Thus, there exists a dire need to protect AMS IP/ICs.

There has been considerable work attempting to protect digital ICs from the above threats, and researchers have recently drawn inspiration from them in order to protect analog ICs. The main strategy in the field of hardware IP protection has been logic locking~\cite{digital_lock3,digital_lock2,digital_lock1}. Logic locking\footnote{Also referred as logic obfuscation and logic encryption in the literature~\cite{book:logic_locking2}.} is a design-for-security (DFS) technique which protects a hardware IP by modifying its design in such as way that it produces erroneous behavior unless properly activated by a key. Recent works which claim to protect AMS and radio frequency (RF) IPs~\cite{AnalogProtection:splitM} against counterfeiting activities include~\cite{rao2017parameter} and~\cite{wang2017thwarting}. Both are performed in a similar manner except for how the key is chosen and applied to the circuit. In the former, the key is used to obfuscate/hide the biasing parameters of the IP and is determined randomly. In the latter, a satisfiability modulo theory (SMT)-solver is used to find ``unique'' key such that all incorrect keys cause the circuit to exhibit behavior far from the intended design. Though the above techniques claim to protect analog IPs by combinational locking, we show in this paper that they are vulnerable to attacks. Specifically, we propose an attack methodology that employs evolutionary algorithms, which is capable of breaking all known analog logic locking techniques. Our main contributions are summarized as follows.
\begin{itemize}
    \item We propose a genetic algorithm (GA) attack to reverse engineer obfuscated analog circuits. Similar to most attacks on digital logic locking, our attack requires the locked netlist and an unlocked IC (i.e., oracle).
    \item We implement the proposed GA attack and demonstrate that it is capable of obtaining both keys and obfuscated parameters of all analog circuit benchmarks locked by SMT combinational locking~\cite{wang2017thwarting} and parametric biasing~\cite{rao2017parameter}. This can be accomplished by GA in 2 minutes in the worst case, and our results show that attack time only increases linearly with key size.
    \item To further demonstrate the efficacy of the GA attack in more complex, real-world scenarios, we apply it on a locked phase-locked loop (PLL) embedded within a  superheterodyne receiver locked by a 512-bit key. Using the output of the receiver instead of the PLL (which would be inaccessible in practice), we are able to extract the 40-bit portion of the key that locks the PLL within 5 minutes.
    \item We compare the proposed GA attack to state-of-the-art attack strategies against analog locking based on SMT solvers, and demonstrate its superiority in extracting keys that more closely match the oracle.
    \item We also explain how GA attack can extend to other analog locking techniques proposed in the literature.
    
\end{itemize}
The rest of the paper is organized as follows. Section~II provides background on analog locking techniques and genetic algorithm. Section~III defines our threat model, describes our proposed attack methodology, and compares it to alternative techniques. Experimental results and discussion are given in Section~IV. Finally, Section~V concludes the paper. 

\section{Background}
\subsection{Related Works on Analog Locking}
\label{subsection:works_analog_lock}
The security of locking or key-based obfuscation often comes down to the size of the key. First and foremost, the key must be long enough to prevent attackers from easily applying brute force. Thus, a 128-bit key will require a circuit to produce $2^{128}$ output functions of which only one should be the correct one. Compared to digital logic locking, it is more difficult to realize such a property in analog circuits since they are designed to work in the presence of process variability~\cite{Polian_AMSsecurity}. Aside from brute force, the key should also be secure from oracle attacks such as SAT~\cite{subramanyan2015evaluating} and SMT~\cite{Azar_Kamali_Homayoun_Sasan_2018}. In oracle attacks, the adversary is assumed to have access to the locked netlist as well as an unlocked IC (i.e., oracle). The locked netlist is in the possession of a third party foundry, but can also be obtained by an end user through reverse engineering. As mentioned above, analog ICs are lower in complexity than digital ICs, and therefore easier to de-process and image to reconstruct the locked netlist. The unlocked chip can be either purchased in a free market or stolen/captured in a controlled supply chain.


\begin{figure}[t]
\subfloat[]{\includegraphics[width=0.24\textwidth,height=0.24\textwidth,keepaspectratio]{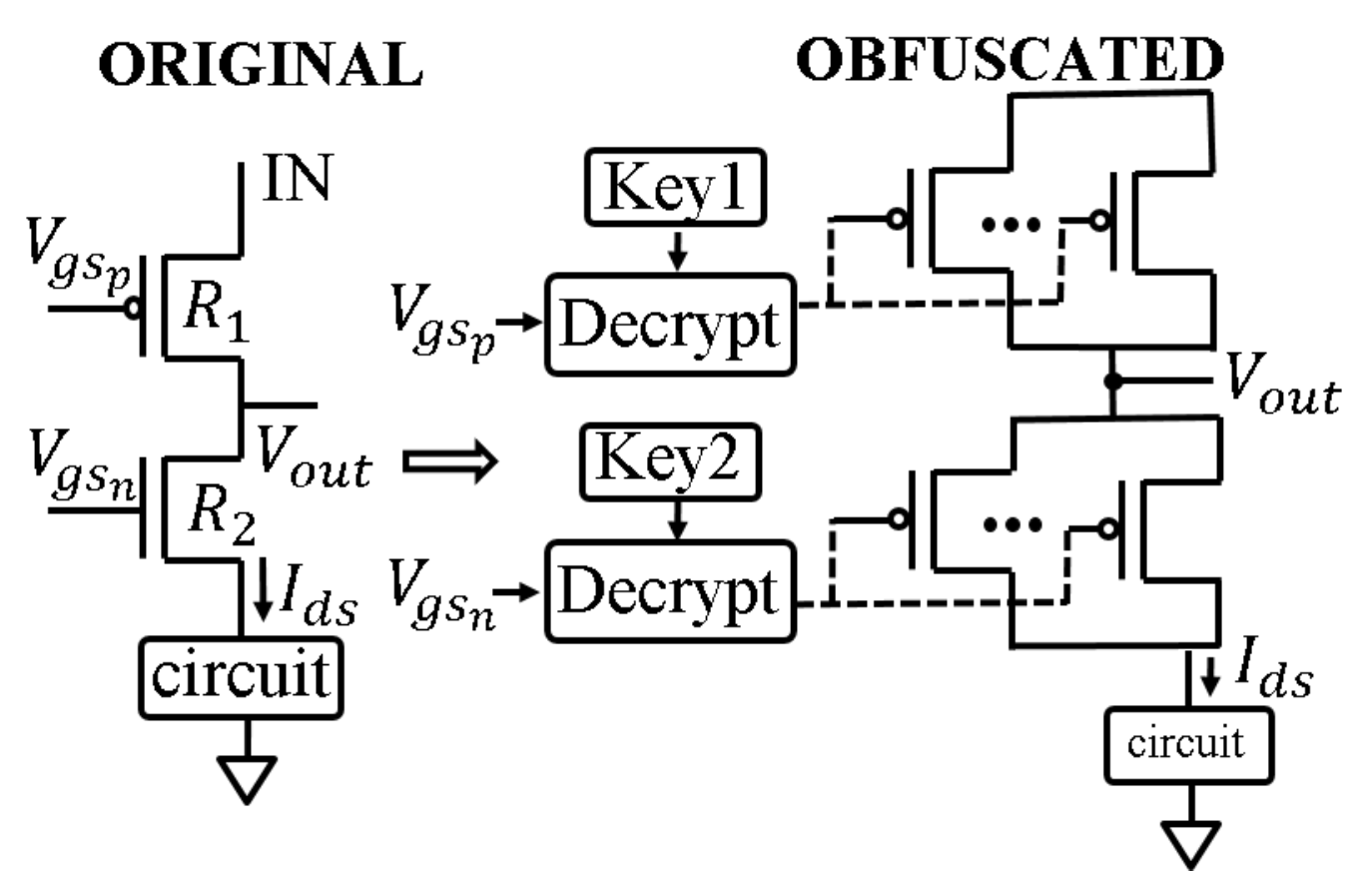}\label{fig:Parameter_biasing}}
\subfloat[]{\includegraphics[width=0.21\textwidth,height=0.24\textwidth,keepaspectratio]{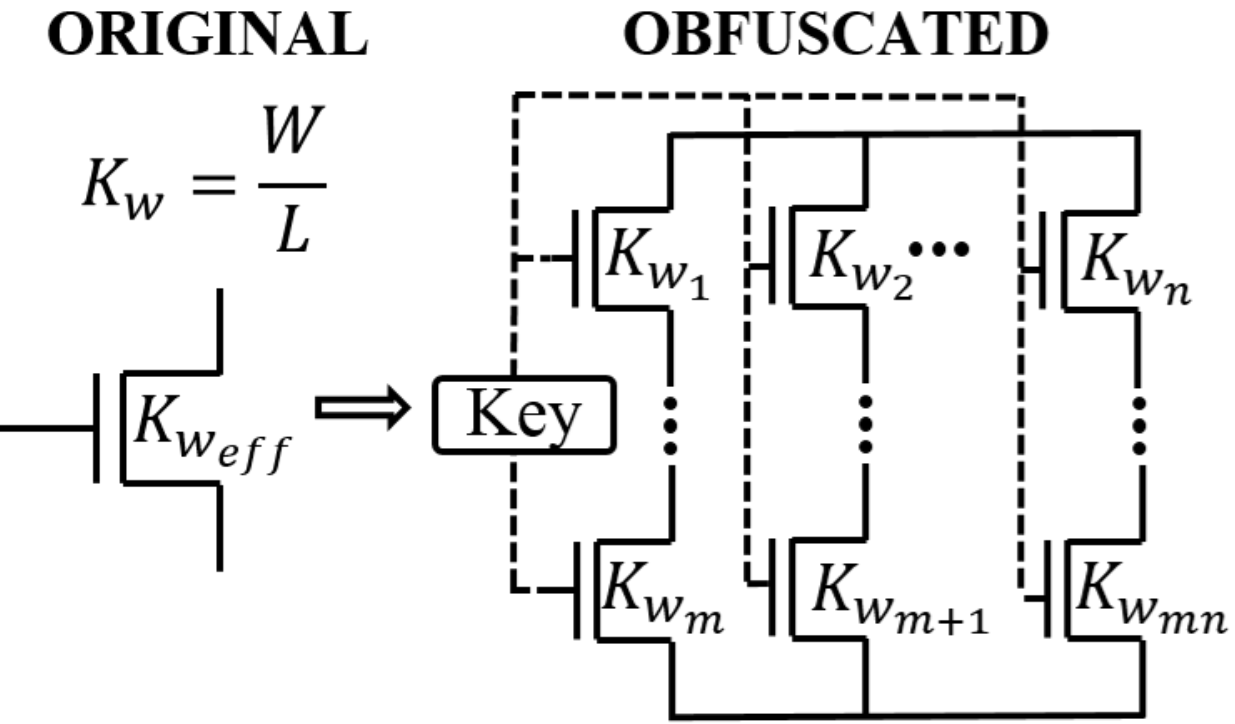}\label{fig:SMT_locking_width}}\\
\subfloat[]{\includegraphics[width=0.235\textwidth,height=0.235\textwidth,keepaspectratio]{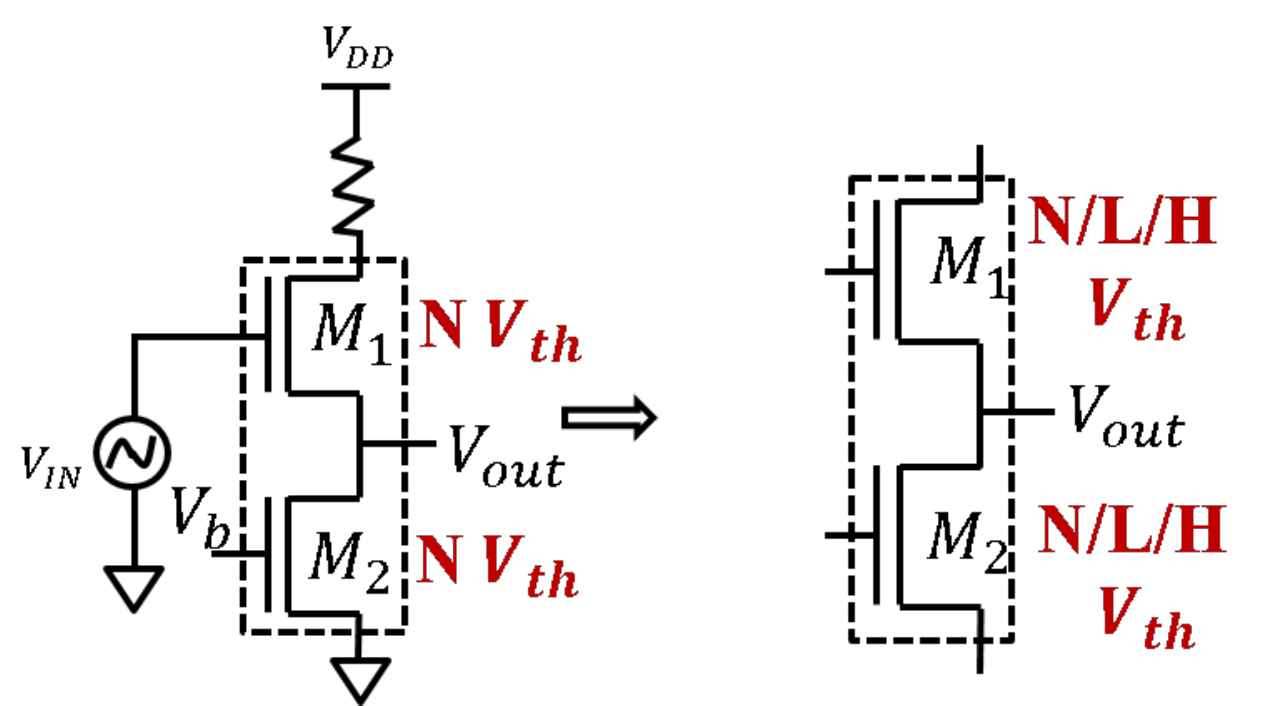}\label{fig:Multithreshold_biasing}}
\hspace{1ex}
\subfloat[]{\includegraphics[width=0.235\textwidth,height=0.235\textwidth,keepaspectratio]{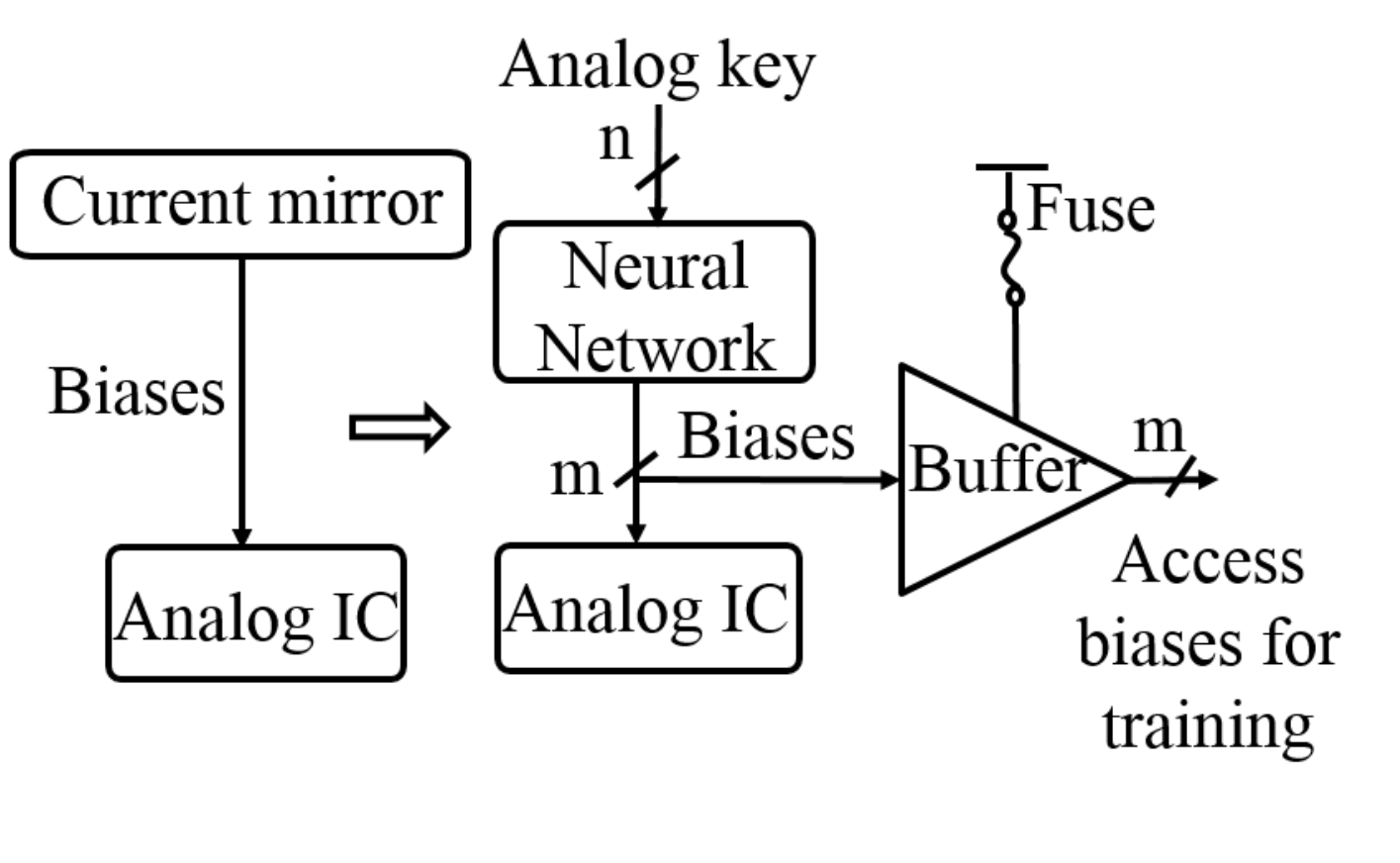}\label{fig:NN_locking}}
\caption{State-of-the-art analog locking techniques. (a) parameter-biasing based locking (PB-Lock)~\cite{rao2017parameter}; (b) SMT-based locking of transistor widths (SMT-Lock)~\cite{raomeshbiasing,wang2017thwarting}; (c) multi-threshold voltage  ($V_\text{th}$) locking~\cite{AnalogProtection:Multithreshold} where N, L and H $V_\text{th}$ refer to normal, low and high $V_\text{th}$ respectively; (d) neural network based locking~\cite{volanis2019analog}.}\label{fig:analog_locking}
\end{figure}

Fig.~\ref{fig:analog_locking} illustrates several of the recent state-of-the-art locking/obfuscation techniques aimed at analog IP/ICs. They can be divided coarsely into the following categories. 

 \vspace{0.5ex}
 
 \noindent\textbf{Parameter-biasing based locking}:
In~\cite{rao2017parameter}, the widths of the transistors are obfuscated using key-based combinational locking of the biasing parameters (current or voltage bias) of the circuit as shown in Fig.~\ref{fig:Parameter_biasing}. However, this technique has one significant drawback which is the fact that it does not ensure the output using the correct key is significantly unique. In other words, there can be multiple keys that provide the correct circuit behavior/performance.

 \vspace{0.5ex}

\noindent\textbf{SMT-based locking/combinational locking}:
To overcome the drawback of parameter-biasing,~\cite{wang2017thwarting} and~\cite{raomeshbiasing} use satisfiability modulo theorem (SMT)-based locking techniques to ensure that only one key provides the correct output while other keys show significant performance deviation. The value of effective width ($W_{eff}$) of the original transistor is obfuscated by replacing it with an $m \times n$ grid of transistors with widths ($W_{mn}$) controlled by a key as shown in Fig.~\ref{fig:SMT_locking_width}.

 \vspace{0.5ex}

\noindent\textbf{Multi-threshold voltage locking}:
The technique described in~\cite{AnalogProtection:Multithreshold} involves using transistors of multiple threshold voltages within the range of $[L V_{th}, H V_{th}]$ to obfuscate actual threshold voltage $N V_{th}$ as shown in Fig.~\ref{fig:Multithreshold_biasing}. This is to ensure significant performance degradation upon application of the incorrect $V_{th}$. This technique might protect against imaging-based reverse engineering since differences in dopants are difficult to distinguish by imaging, but will not protect against overproduction as foundry knows the correct $V_{th}$ during manufacturing.

 \vspace{0.5ex}

\noindent\textbf{Neural-network based locking}:
In an attempt to counter model approximation attacks, on-die analog neural network was recently proposed in~\cite{volanis2019analog} as shown in Fig.~\ref{fig:NN_locking} such that the trained network serves as a lock and its analog input serves as the key. Only a correct pair of key and lock will set the correct bias for the analog circuit. This is to ensure that the keys are only used in the analog domain and that the number of possible keys is increased, consequently impairing the effectiveness of the model approximation attacks~\cite{volanis2019analog}. 
Although this can be thought of as a promising solution against key guessing and brute force attacks, their neural network-enabled method is still vulnerable to attacks, ranging from reverse engineering to ones which are agnostic with regard to the key domain.

To recap, we stress that while the above-mentioned methods offer a few adaptable solutions for analog design obfuscation, they are susceptible to either reverse engineering or lock removal attacks~\cite{Torrance_RE}. More importantly, the relationship between the key and the performance of the IC is simple enough to be approximated, and therefore, the model underlying the obfuscation can be learned.
 
In this paper, we demonstrate such an attack and show that this relationship can be easily exploited to guide a genetic algorithm (GA), and consequently, unlock the design.  
In parallel to our work, when our paper was under review, a GA-based attack against logic locking in a digital circuit has been proposed~\cite{genunlock}. 
Needless to say, due to clear differences between analog and digital locking schemes cf.~\cite{Polian_AMSsecurity}, this attack cannot be compared to ours directly. 
In this regard, we will show how to launch an oracle attack on a locked analog circuit to extract either the correct key \textit{or} the hidden internal parameters.

\subsection{Introduction to Genetic Algorithm}
\subsubsection{Basic Idea and Operation} 
\begin{figure}[t]
\centerline{\includegraphics[width=0.45\textwidth,height=0.45\textwidth,keepaspectratio]{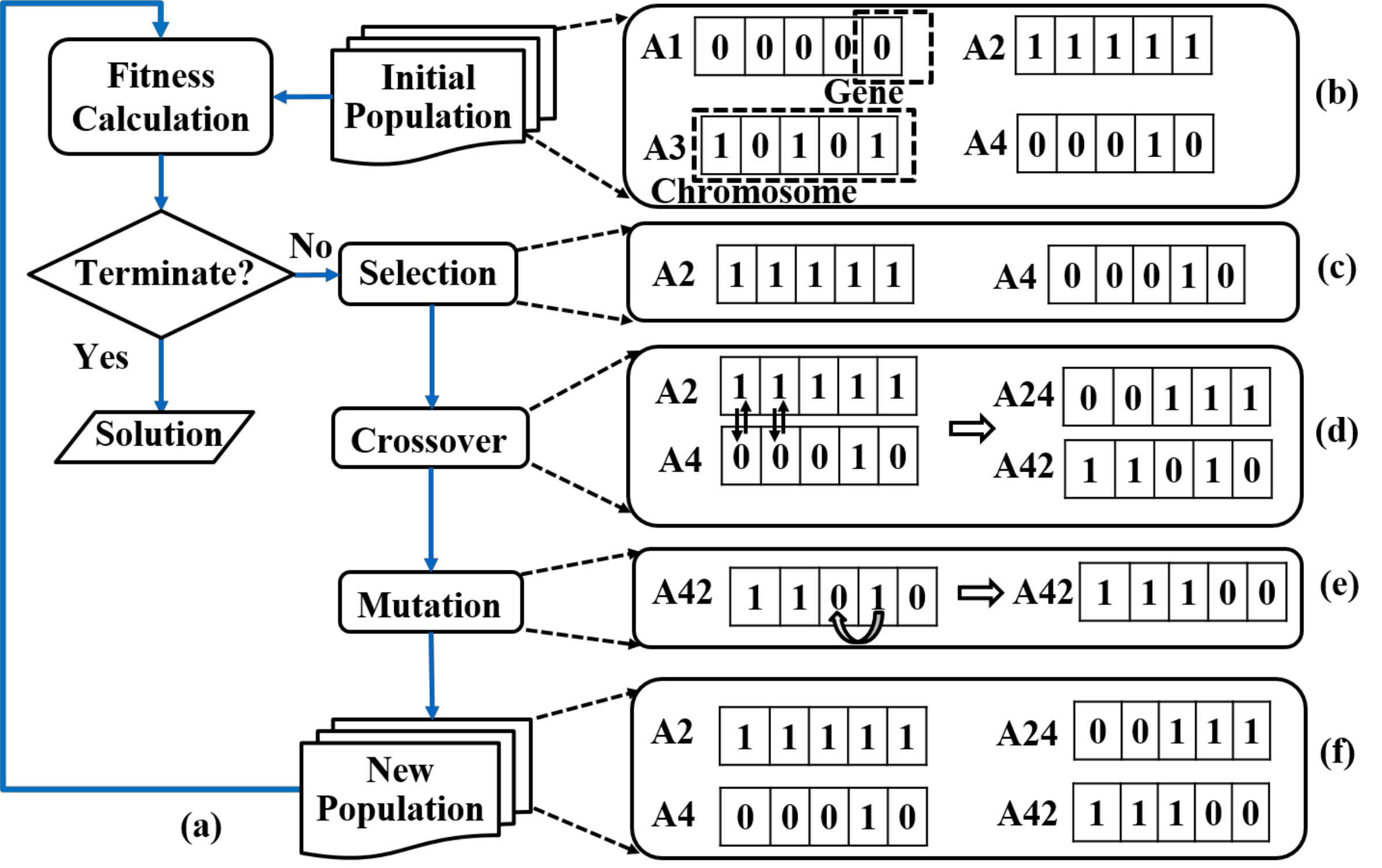}}
\caption{(a) Flowchart of the genetic algorithm (GA); Example showing (b) population initialization; (c) selection of fit members from the population; (d) crossover between fit members to create new members; (e) mutation of genes in certain members; (f) the new population after selection, crossover, and mutation.}
\label{fig:GA_intro}
\end{figure}

The genetic algorithm (GA) is an optimization algorithm inspired by the process of natural selection. To this end, GA applies three evolutionary operators -- selection, crossover, and mutation -- to a population of \textit{chromosomes} as shown in Fig.~\ref{fig:GA_intro}. Chromosomes are created by encoding potential solutions to the problem as strings of real numbers or binary bits (e.g., A1 through A4 represents different chromosomes in Fig.~\ref{fig:GA_intro}b). A real number or a bit is referred to as a \textit{gene} of the chromosome. Representing a candidate solution or genetic encoding is problem dependent and one of the major components of the GA~\cite{intro_ga}.
After encoding, a large population of random chromosomes, each of $L$ genes, is created. Each chromosome is then tested to see how suitable it is at solving the problem at hand. This is accomplished by using a \textit{fitness function} which assigns a \textit{fitness value} to each solution. Depending on the fitness value, only a few chromosomes are selected for the next stage (see Fig.~\ref{fig:GA_intro}c). New population members are created by merging two previously fit chromosomes in a process called \textit{crossover} (see Fig.~\ref{fig:GA_intro}d) and mutating certain genes of the resulting chromosome in a process called \textit{mutation} (see Fig.~\ref{fig:GA_intro}e). These processes are controlled by crossover and mutation rates, which determine the probabilities of performing crossover and mutation of each gene. In a simple GA, only a fraction of the individuals is replaced in each generation, and the selection process is biased towards highly fit individuals. This \textit{evolution} of individuals is repeated until a stopping criterion is met -- either a solution of desired fitness is found or a limit on the number of iterations is reached. 

\label{subsection:GA_intro}

\subsubsection{Complexity and Effectiveness of GA} GA is a highly non-linear search algorithm, and this inherent feature makes it difficult to model its behavior. 
The \textit{schema theory}, also called fundamental theorem, is one of the accepted models which helps to understand the way GA works. 
In this regard, a schema is defined as a specific pattern describing a set of chromosomes. 
For instance, in Fig.~\ref{fig:GA_intro}b, the chromosomes A1 and A4 belong to the schema $H=(0\; 0\; 0\; *\; 0)$. 
In this model, the symbol $*$ shows that only the gene at the fourth position can have a value of either 1 or 0, and other positions are fixed, i.e., have a value of 0. 
The \textit{order} of a schema $o(H)$ is the number of its fixed positions, e.g., $o(H)=4$ since there are four positions with zero in common between A1 and A4. 
In addition, $\rho(H)$ is referred to as the \textit{defining length} of the schema, that is the distance between its first and last specific positions. 
Moreover, $m(H,t)$ denotes the number of individuals belonging to the schema $H$ at a particular generation $t$. 
Accordingly, the fitness of the schema $F(H)$ can be defined as the average fitness of all strings belonging to the schema. 
We can now define the schema theorem as 
\begin{equation}
\small
m(H,t+1)\geq\dfrac{m(H,t)F(H)}{F_m} \Big(1 - p_c\dfrac{\rho(H)}{L-1}\Big)\big(1-o(H) p_m\big), 
\label{eq:Holland_schema}
\end{equation}
where $F_m$ is the average fitness of the population at generation $t$, $p_c$ is the crossover rate, and $p_m$ is the mutation rate.
The most important interpretation of this theorem is that a pattern describing a set of chromosomes will remain or disappear in the next generations. 
More importantly, this theorem is a formalization of how the quality of the next generation can be predicted. 
In other words, it identifies the ``building blocks'' of a good solution, which are short, low-order, and high-performance schemata. 
Furthermore, it relates this quality to the crossover, believed to be a major source of the creative power of GAs~\cite{mitchell1995genetic}.  

\subsubsection{Applications of GAs} GAs are used in a variety of applications that involve solving complex problems such as machine learning, automatic programming, and data modeling. They have great advantages compared to conventional optimization methods such as calculus-based optimization methods (hill-climbing and gradient-ascent), exhaustive search methods, and random search methods~\cite{book:GA_intro}. Unlike these methods, GAs are more likely to find a global optimum instead of getting stuck in a local optimum. Further, GAs do not require extra information like derivatives or gradients. The only mechanism that guides their search is the numerical fitness value of the candidate solutions. This allows them to function when the search space is noisy, nonlinear, and derivatives do not exist. GA is also more flexible and can be tuned depending on whether accuracy or efficiency is more important. Thus, GA can apply to many more situations than traditional algorithms.

\subsubsection{GA in Circuit Design}
\label{subsection: GA_circuit_design}
Among various applications, evolutionary computation has been demonstrated as promising and powerful for computer-aided design (CAD) of electronic circuits.
For this purpose, the challenging task of satisfying multiple objectives (e.g., cost, power consumption, performance, and reliability) and selecting the most appropriate topology, i.e., optimization of the design, can be accomplished through evolutionary algorithms, and more specifically, GAs~\cite{zebulum2018evolutionary}.
In particular, GAs have been widely adopted to handle the problems related to not only design optimization, but also the synthesis of analog circuits. 
The popularity of GAs stems from their inherent ability to deal with optimization and search problems, where the size or complexity of the problems makes the application of other optimization techniques infeasible. 
Moreover, and more importantly for our approach, GAs require no additional information on the search space. 
Nevertheless, in this paper, we focus on another aspect of designing a circuit -- the security of the design -- which can also be tackled by GAs successfully. 
Ironically enough, to assess the security of analog circuits, we launch GA-based attacks.

\section{Attack Approach}
\subsection{Threat Model}
\label{section:threat_model}
We assume a threat model consisting of either an untrusted foundry or an end user who possesses the necessary tools and skills needed to reverse engineer and counterfeit the circuit from a GDSII file~\cite{Torrance_RE}. In addition, the circuit is assumed trusted and devoid of any malicious components. Below we summarize the assumptions made by our proposed attack.
\begin{itemize}
    \item {The attacker has access to an unlocked IC (oracle) and can measure correct input/output signals from it.}
    \item{The attacker has the obfuscated/locked netlist.}
    \item{The attacker does not need to have knowledge of the locking algorithm used, but he or she needs to know where in the IP/IC that the key is being applied.}
\end{itemize}

\subsection{Proposed Attack Methodology}
\label{section:GA_attack_method}

\begin{figure}[t]
\centerline{\includegraphics[width=0.35\textwidth,height=0.35\textwidth,keepaspectratio]{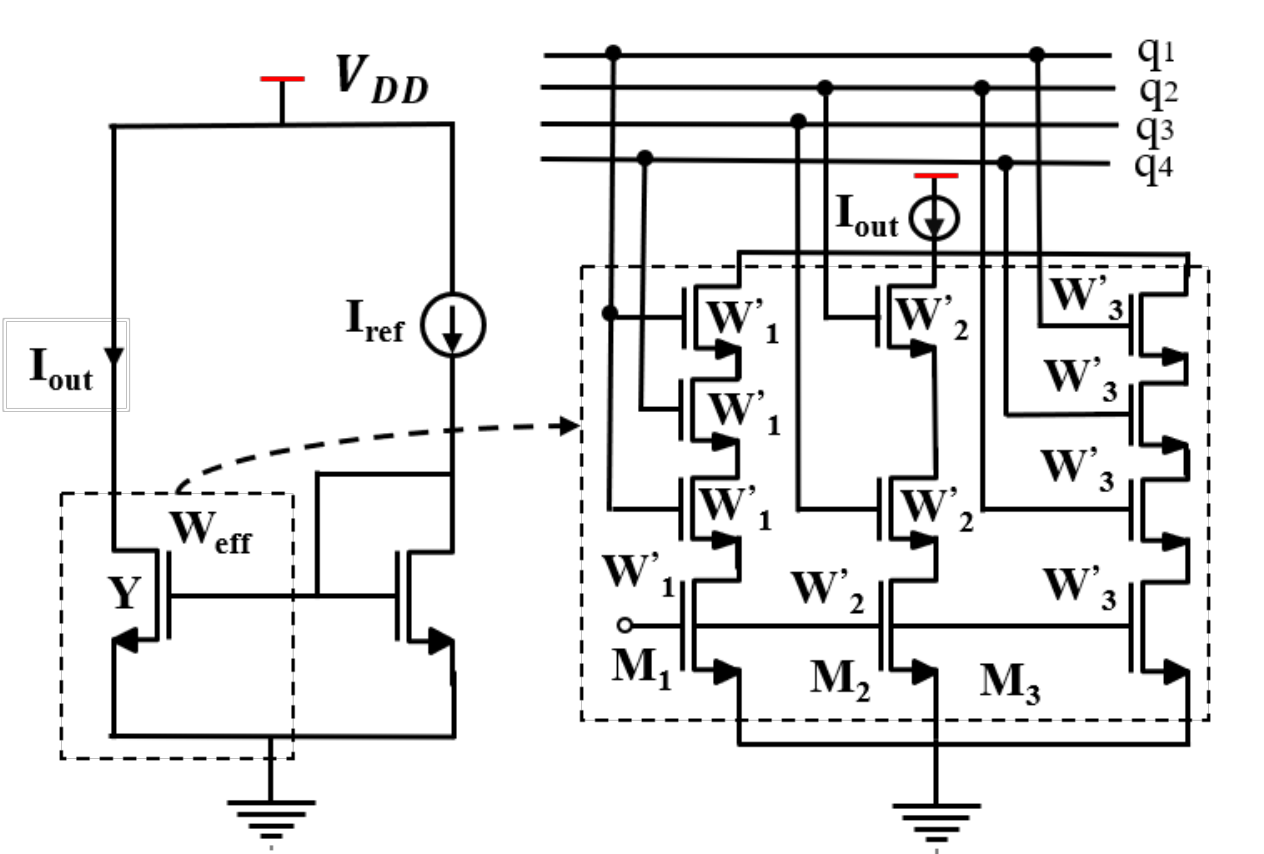}}
\caption{Combinational locking example. (a) Original current mirror with the transistor $Y$ of width $W_{eff}$. (b) Additional transistors connected to key bits to replace $Y$. $q_1$, $q_2$, etc. represent key bits.}
\label{fig:Current_mirror}
\end{figure}

\begin{figure}
    \centering
    \subfloat[]{
   \includegraphics[width=0.20\textwidth,height=0.20\textwidth,keepaspectratio]{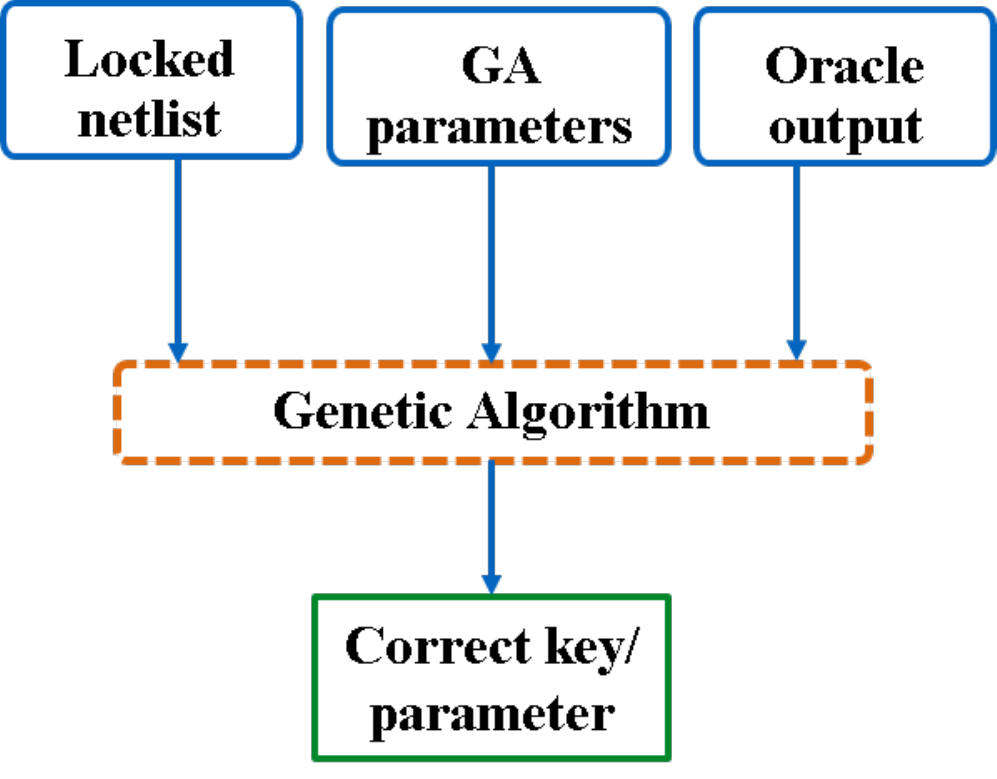}\label{fig:GA_attack}}
   \hspace{5pt}
      \subfloat[]{
   \includegraphics[width=0.20\textwidth,height=0.20\textwidth,keepaspectratio]{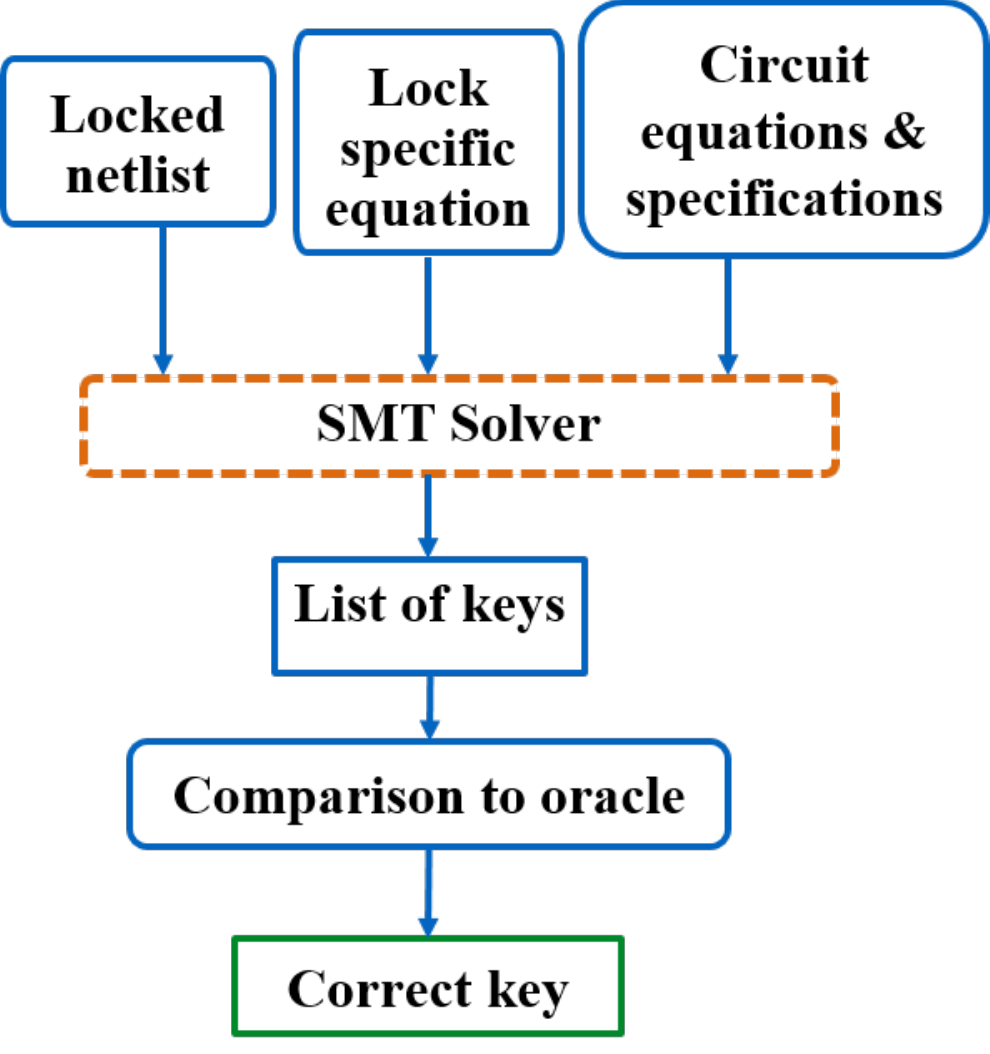}\label{fig:SMT_attack}}
    \caption{Comparison between two attack methods. (a) Attack process using GA; (b) Attack process using SMT-based method. }\label{fig:Attack_methods} 
\end{figure}

As a running example, we consider combinational locking techniques where analog circuit components are controlled by key bits as shown in Fig.~\ref{fig:Current_mirror}. This obfuscates the correct value of the circuit parameters such as the width of the transistor Y ($W_{eff}$) in the original design. Only on application of the correct key, the circuit operates correctly. As shown in Fig.~\ref{fig:GA_attack}, our attack takes a locked netlist and an oracle output as inputs.  Then, based on the encoding of the problem, GA will either evolve various combinations of key bits or various values of obfuscated circuit parameters (such as Y's width in Fig.~\ref{fig:Current_mirror}) to obtain an output that matches the oracle's.

Before running the algorithm, parameters as described in Section~\ref{subsection:GA_intro} should be set by the user. These parameters, mainly the number of chromosomes or solutions ($N$), fitness function, selection criterion, crossover rate, mutation rate, and stopping criterion, guide the algorithm to reach or converge towards a solution. Below we will describe these parameters and a general procedure in extracting the correct key/parameters from a locked circuit. 
\begin{enumerate}
    \item To start the algorithm, the first \textit{generation} of candidate solutions is created by generating $N$ random chromosomes. $N$ is usually chosen small so that the algorithm can converge towards a solution quickly. If one is evolving the circuit parameters, each chromosome will be \textit{encoded} with real numbers. Alternatively, if one is evolving the key, the chromosome encoding is binary. The length of the chromosome ($L$) is equal to the key size $k$ in this paper.
    
    \textbf{Example:} For the circuit locked as shown in Fig.~\ref{fig:Current_mirror} with a 4-bit key, a potential chromosome will either be $[1010]$ or [$480.3nm$] if we are evolving key bits or Y's width value, respectively.
    \item Next, one defines the fitness function, i.e., an objective function that estimates how closely a given design solution (chromosome) solves the problem. For an analog IC, there could be many important specification parameters of the circuit and its output (e.g., throughput, linearity, bandwidth, gain, etc.) used by the fitness function. 
    For the benchmarks in later sections, we will use either the output frequency response or the output transient response of the circuit as parameters of our fitness function. A general fitness function $F$ which helps guide the GA towards the correct solution is~\cite{murata1995moga}
    \begin{equation}
    \small
    F:= \sum_{j=1}^{m}\sum_{i=1}^{n}(f_{i,j})^2, \text{s.t }
    f_{i,j} = E_{i,j}-O_{i,j}, \label{eq:fitness}
    \end{equation}
    where $f_{i,j}$ is an individual fitness criterion applied to a problem with $n$ data points in our multi-objective fitness function and $m$ is the number of fitness criteria used. 
    For each criterion $f_{i,j}$, $E_{i,j}$ is the target output obtained from the unlocked IC for parameter $i$ and the $j^{th}$ fitness criteria, and $O_{i,j}$ is the obtained output after simulating the netlist with the chromosome solution. 
    
    \textbf{Example: } 
    In most of our later experiments, $m=1$ and the fitness criterion corresponds to output frequency or transient response. $i$ denotes the index in an output data vector collected over frequency or over time.
    
    \item The best-fit chromosomes (i.e., those with the smallest value of $F$) are selected for the next steps of the evolution while the rest are discarded from the population. In this paper, we employed the \textit{roulette wheel selection}~\cite{intro_ga}, in which the slots of a roulette wheel are sized according to the $F$ of each chromosome solution in the population, and a chromosome is selected by spinning the roulette wheel. 
    \item The selected chromosomes are used to create new chromosomes or offspring using the crossover operator. The crossover operator swaps a part of the sequence of two of the selected chromosomes to create two offspring as discussed in Section~\ref{subsection:GA_intro}. Note that the crossover operator is only used for binary-encoded chromosomes in this paper.
    \item These new chromosomes are mutated with a certain rate to determine how many bits in the offspring will be flipped from 0 to 1 and vice versa. For chromosomes coded with real numbers, the value of the chromosome will change by the mutation rate.
    \item Steps 2 to 5 are repeated until a stopping criterion is met. Each iteration is called a \textit{run}, and at the end of each run, there is usually at least one chromosome with the smallest fitness value. This could be the final solution or additional runs may be needed to converge towards the fittest individual. The number of runs is one of the stopping criteria for halting the optimization process. Similarly, time allocated for each run and total time to run the algorithm can also be employed. Depending on the progress made, one can change these parameters including the mutation rate, the selection criteria, the crossover point, etc. for better optimization.
\end{enumerate}
\subsection{Two-Pass Variant of Proposed Methodology.} \label{sec:2pass}
In the case of attacking complex benchmark circuits locked with binary key of length $k$, the search space for GA is $2^k$. One way to reduce this space is to apply the GA attack twice. In the first pass, the chromosomes can be encoded using real numbers in order to identify the value of obfuscated circuit parameters. In the second pass, the result from the first pass can be used an additional fitness criteria to guide the GA when it searches for the correct key. In cases where the obfuscated parameters (e.g., transistor widths and lengths, threshold voltages, etc.) take on discrete values, the search space for the two-pass approach will effectively reduce from $2^k$ to linear in the discrete values of the obfuscated parameters. 

\subsection{Comparison to Other Attack Techniques}
As it was previously demonstrated that SMT solvers can successfully de-obfuscate locked digital circuits~\cite{Azar_Kamali_Homayoun_Sasan_2018}, it may seem tempting to apply the same method against obfuscated analog circuits. 
This section aims to show how this could be possible and highlights the limitations of such an attack.
\subsubsection{SMT-based Attack on Combinational Locking}
\label{subsection:SMT_attack}
Attacking circuits locked using combinational techniques~\cite{rao2017parameter,raomeshbiasing,wang2017thwarting} involves formulating a system of equations that mirror how the locking has been implemented. That means to launch an SMT-based attack, the adversary needs to have detailed knowledge regarding the locking technique unlike launching an attack based on the proposed GA. The procedure taken by SMT-based attack to break the defense techniques in~\cite{wang2017thwarting,raomeshbiasing,rao2017parameter} is shown in Fig.~\ref{fig:SMT_attack} and is briefly explained below.
\begin{enumerate}
    \item {Identify the portion of the circuit where the lock has been implemented and then figure out the type of obfuscation used, e.g., parameter-biasing~\cite{rao2017parameter} or SMT-based~\cite{wang2017thwarting,raomeshbiasing}.}
    \item {Find the obfuscated circuit parameters such as the width of transistors and their dependency on the key inputs based on the locking technique and circuit specifications.\\
    \textbf{Example.} Transistor Y of width $W_{eff}$ is replaced by $m\times n$ transistors where transistors in the same column $i$ has same width value $W'_i$. That means $W_{eff}$ is replaced by $W'_i = \{W'_1, W'_2,....,W'_n\}$,  where $W'_i$ is connected to $m$ stacked transistors of the same width controlled by the key vector $q'$ which represents multiple bits of the key $q$. This in turn obfuscates the output current
    \begin{equation}
    \small
    I_{out}=\Big(\sum_{i=1}^{n} W'_i \prod_{j=2}^{m}q'_{ij}\Big) I_{ref},\label{eq:lock}
    \end{equation}
    where $q'_{ij}$ represents the value of key bit connected to the transistor at row $i$ and column $j$ and is $1$ if no transistor exists at that location~\cite{wang2017thwarting}. $I_\text{ref}$ is part of the circuit specifications and needs to be known by the attacker.}
	\item {Formulate theoretical circuit equations for the obfuscated bias based on the benchmark.}
	\item {Equate the equations from the first and the second step and use the solver to find a key or list of keys. If a list is returned, compare the output obtained from the keys with the oracle to determine the correct key.}
\end{enumerate}

\subsubsection{SMT Solvers -- Challenges and Possibilities}
\label{subsection:SMT_challenges}
Similar to GAs, SMT solvers first found application in design and verification of analog and mixed-signal (AMS) circuits, e.g.,~\cite{walter2007bounded,saif2016pareto}. 
In addition to verification, SMT solvers have been utilized to enhance understanding of the functionality of small components and sub-circuits in a larger system. 
Although challenging to determine the mapping between the inputs and outputs of sub-circuits with equivalent functionality, it has been demonstrated that SMT solvers can solve this problem~\cite{gascon2014template}, cf.~\cite{keshavarz2018survey}. 
Recently, SMT solvers were used to assess the security of hardware primitives, namely by launching attacks against obfuscated digital circuits\footnote{During the time that we were developing the ideas presented here, an SMT-supported attack against analog locking was proposed in~\cite{jayasankaran2019breaking}. 
While we did not have access to~\cite{jayasankaran2019breaking} and could not elaborate on its details, we expected the main differences to be similar as between our approach and~\cite{Azar_Kamali_Homayoun_Sasan_2018}. 
For the final version of our paper, as~\cite{jayasankaran2019breaking} is now available, this has been examined and found to be true. }~\cite{Azar_Kamali_Homayoun_Sasan_2018}. 
It has been claimed that SMT solvers can overcome obstacles that the attacker has to face by solving problems classified as constraint satisfaction problems (CSPs). 
Formally, such a problem is defined by a set of variables $X_1,\cdots, X_n$, where each variable (e.g., $X_i$) has a nonempty domain $D_i$ ($1\leq i \leq n$) containing possible values that can be assigned to $X_i$. 
Along with the set of variables, a set of constraints $C_1,\cdots, C_m$ is defined, with $C_i$ specifying the restricting rules on the values that the variables can simultaneously take. 
When finding a solution to a CSP, the solver must find an assignment of the values to variables so that the values are chosen from the respective domains, and each and every constraint is satisfied (for more details, see~\cite{barrett2018satisfiability}). 
The solver can deliver either a set of possible solutions or one solution, with no preference. 
Nowadays, a typical SMT solver is composed of a SAT solver, crucial for solving Boolean satisfiability problem, as well as theory solvers for decidable theories in concrete domains, e.g., linear integer arithmetic, arrays and bit-vectors, etc.\footnote{Note that our main focus lies on the so-called lazy approach, in which theory solvers are involved in the problem-solving process. This is due to their superior efficiency in the context of this study. }~\cite{de2013strategy}. 

With regard to this definition, it is tempting to formulate the problem of finding the correct key of an obfuscated analog circuit as a CSP problem, and accordingly, apply an off-the-shelf SMT solver to unlock the circuit. 
However, one has to be careful about the challenges and limitations that are inevitably involved in SMT solvers. 
First and foremost, an SMT solver is a collection of heuristics used to combine several algorithmic proof methods. 
A theoretical framework, namely Davis–Putnam–Logemann–Loveland (DPLL) scheme, must have been principally applied to integrate the theory solvers together with a SAT solver. 
Nevertheless, in typical SMT solvers, the DPLL scheme has been either substituted by or combined with a set of heuristics~\cite{de2013strategy}. 
Consequently, from one SMT to another, it may be impossible to achieve the same result for given a problem. 
More importantly, due to a lack of detailed knowledge of how the heuristics can be controlled, it may be possible to neither generalize the results to similar systems under test (e.g., similarly obfuscated circuits) nor repeat the same experiment successfully. As a prime example, the formula preprocessing, i.e., the process of translating the given problem into a formula understandable by the solver, can be handled by different SMT solvers in various ways. 
In the literature, this problem is well formulated as the ``strategic control'', meaning that end-users are given theoretical and practical tools and methods to control core heuristic aspects of the SMT. 
In this regard, a strategy, i.e., general search mechanisms, should be adopted to reduce the search space by focusing on a particular class of problems~\cite{de2013strategy}. 
Interestingly enough, to devise such effective strategies, evolutionary algorithms, particularly, GAs have been used to improve SMT solver efficiency, e.g.,~\cite{ramirez2016towards,zhang2006extracting,ramirez2018framework}. 

Moreover, even if heuristics mentioned above could be tailored to the needs of particular known classes of problems, see, e.g.,~\cite{smtlib}, they can perform poorly on other classes.  
One may argue that in the context of obfuscated circuits, in general, the netlist of circuits can be represented by graphs, and then first-order theories involved in SMT solvers can be applied to tackle the problem of de-obfuscation.  
Hence, it might be concluded that the design of SMT solvers is appropriate for these problems. 
It is clear that even in this case, SMT-based approaches suffer from two main shortcomings: 
(1) the preprocessing step (i.e., translation of the netlist to the graph) increases the computational overhead, and (2) to obtain the graph representing a given netlist, a detailed specification of the circuit is required. 
The latter problem may seem relatively simple, when considering digital circuits; however, for analog circuits, this step involves a precise specification of the circuit. For example, a \textit{current internal to the chip}, $I_\text{ref}$, is needed to formulate the equations and inequalities that describe the circuit shown in Fig.~\ref{fig:Current_mirror}. This quantity is unlikely to be available in practice. 
This is in contrast to GA, where the inputs fed into the algorithms are \emph{solely} the locked netlist and the outputs collected from the locked circuit (see Fig.~\ref{fig:Attack_methods}). 
Comparing the steps of the GA and SMT-based attacks makes this further evident. 
More specifically, for the GA-based attack, the above point refers to the first step (i.e., generating the chromosomes), whereas steps 1-2 in the case of the SMT-based attack should be taken. 
These steps 1-2, as discussed before, are much more complicated as they are not only locking-technique dependent, but also circuit-specification dependent.

\section{Experimental Results and Discussions}
\subsection{Experimental Setup}
\label{section:experiment_setup}
\begin{figure}[t]
\centerline{\includegraphics[width=0.42\textwidth,height=0.42\textwidth,keepaspectratio]{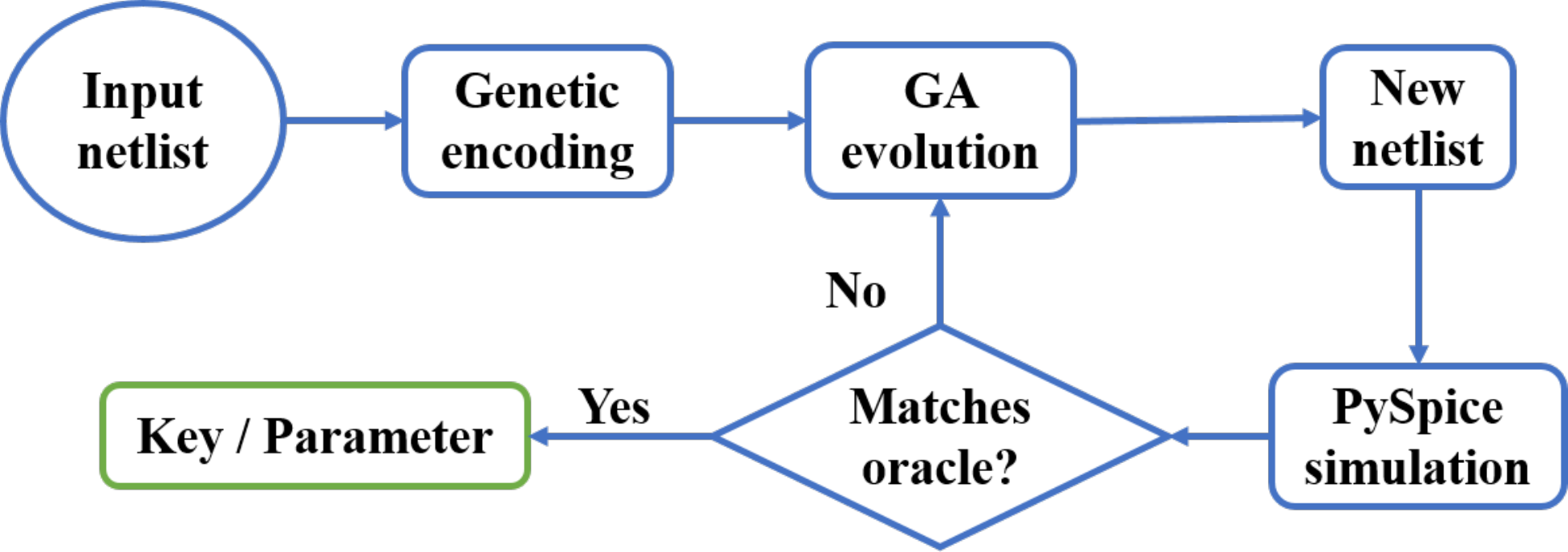}}
\caption{Experimental setup showing the procedure used in extracting the obfuscated parameters and/or keys using GA.}
\label{fig:GA_experiment_setup}
\end{figure}

The proposed GA attack is validated primarily on four different circuit benchmarks based on 180~nm generic process design kit (GPDK) technology: operational transconductance amplifier (OTA), fourth-order Gm-C band-pass filter (BPF), phase-locked loop (PLL), and a triangular waveform generator (TWG) used in class-D amplifiers. We also perform an additional experiment on a superheterodyne receiver. All experiments are run on a CPU which has a x64-based Intel Xeon 3.3~GHz Processor and 32~GB of RAM. The steps taken to carry out the proposed attack are shown in Fig.~\ref{fig:GA_experiment_setup}. The circuits are designed and obfuscated with a $k$-bit key where $k=16, 32, 40,$ and $64$ using SMT-based combinational locking~\cite{wang2017thwarting} and parameter-biasing locking~\cite{rao2017parameter}, henceforth referred to as SMT-Lock and PB-Lock respectively\footnote{The locking techniques are chosen to illustrate the proposed approach and provide a comparison to SMT-based attacks. However, the GA attack should generalize to other locking schemes such as~\cite{AnalogProtection:Multithreshold,volanis2019analog} (see Section~\ref{sec_takeway}).}. The correct output of the unlocked circuit is recorded from the design's output pins as well. The attacks are carried out using GA as described in Section~\ref{section:GA_attack_method}. The creation of chromosomes, the genetic evolution, and the corresponding simulations of the evolved netlist are entirely done within the Python environment. We use PySPICE which is an open-source module that can simulate and manipulate SPICE netlists in Python by interfacing Python to the Ngspice simulators~\cite{PySpice}.


\noindent\textbf{GA Attack Settings.} As discussed in Section~\ref{section:GA_attack_method}, we need to first set the GA parameters to attack the circuit benchmarks.  $N$ is set to $40$. Roulette wheel selection is used as the selection criteria. For binary-encoded chromosomes, single-point topology aware crossover is chosen as the crossover operator where bits to right of the crossover point (half-way point of the key size in our method) are swapped between the two parent chromosomes~\cite{intro_ga}. Mutation rate is chosen to be $5\%$. The obfuscated netlist combined with the generated chromosome is then evaluated for fitness. In general, either output frequency response or transient response is used as a fitness criteria. In other words, the number of fitness criteria is one ($m=1$), in which case Equation~\eqref{eq:fitness} simplifies to 
\begin{equation}
\small
F=\sum_{i=1}^{n}(E_i - O_i)^2.\label{eq:fitness_simplified}
\end{equation}
In this paper, we do not halt the algorithm until we get the desired output. For example, if the frequency response is used as a fitness criterion and the frequency response of the evolved circuit ($O_i$) is equal to the actual output of the unlocked circuit ($E_i$), then the GA halts. Further, if the best fitness value among the chromosomes does not keep improving up until a certain generation, we perform survival selection with the \textit{Age-fitness Pareto algorithm}~\cite{zebulum2018evolutionary}. Survival selection culls the combined population of parents and offspring using tournament selection with replacement. 
This process was implemented with a Pareto tournament scheme in which two random members of the combined population were selected~\cite{zebulum2018evolutionary}. If one of the pair had both lower fitness and higher age than the other, it was thrown out. The survivor was then returned to the pool. This continued until the population size was reduced back to $40$.
Age was defined as the number of generations in which an individual had been present in the population. Offspring inherited the age of the older parent in the case of crossover. One new, randomly generated individual with an age of $0$ and created in the same way as members of the initial population was added to the population in each generation.

\noindent\textbf{SMT Attack.} Similarly, an attack is also carried out using the SMT-based method discussed in Section~\ref{subsection:SMT_attack}. Z3 solver~\cite{Z3_solver}, an open-source SMT-solver is used for this purpose. Circuit equations, equations based on the locking technique, and the circuit specifications were used to extract a list of candidate keys. 

In our experiments, we consider two scenarios:

\vspace{0.5ex}

\noindent\textbf{Case 1: GA Attack to Find Obfuscated Parameters.} Locking techniques in~\cite{rao2017parameter,wang2017thwarting} are both forms of combinational locking where an equivalent width ($W_{eff}$) of the current or voltage biasing circuit is obfuscated using the key. In this case, we use GA to calculate $W_{eff}$ by encoding each chromosome with real numbers and evolving them (different width values) to match the actual circuit output. Essentially, the result is a circuit design with the lock removed and a specification that matches the unlocked chip. In this case, the search space for the GA reduces from exponential ($2^{k}$) to linear (i.e., discrete width values of the transistor).

\noindent\textbf{Example.} Suppose the IP to be locked is a current mirror, $Y$ shown in Fig.~\ref{fig:Current_mirror}, and the attacker has access to the output pin meaning, in this case, he or she access to the $I_{out}$. To protect this IP, the bias transistor is split into multiple branches where each branch $i$ contains additional transistors of sizes $W^{'}_i$ that are controlled by the key. Using $I_{out}$ as a fitness criteria for our GA, we calculate $W_{eff}$ without considering the additional transistors used. The discovered width can also be used as an additional fitness criterion to extract the correct key from the locked circuit for Case 2.


\vspace{0.5ex}

\noindent\textbf{Case 2: GA and SMT Attacks to Find Obfuscation Key.} The circuits locked with SMT-Lock and PB-Lock are attacked using GA by applying multiple-fitness criteria including the equivalent $W$ from the results of Case 1. In addition, we employ an SMT-solver to find keys for comparative purposes.
\subsection{Results of Case 1}
\label{section:Case1_results}


\begin{figure}[t]
\centerline{\includegraphics[width=0.45\textwidth,height=0.45\textwidth,keepaspectratio]{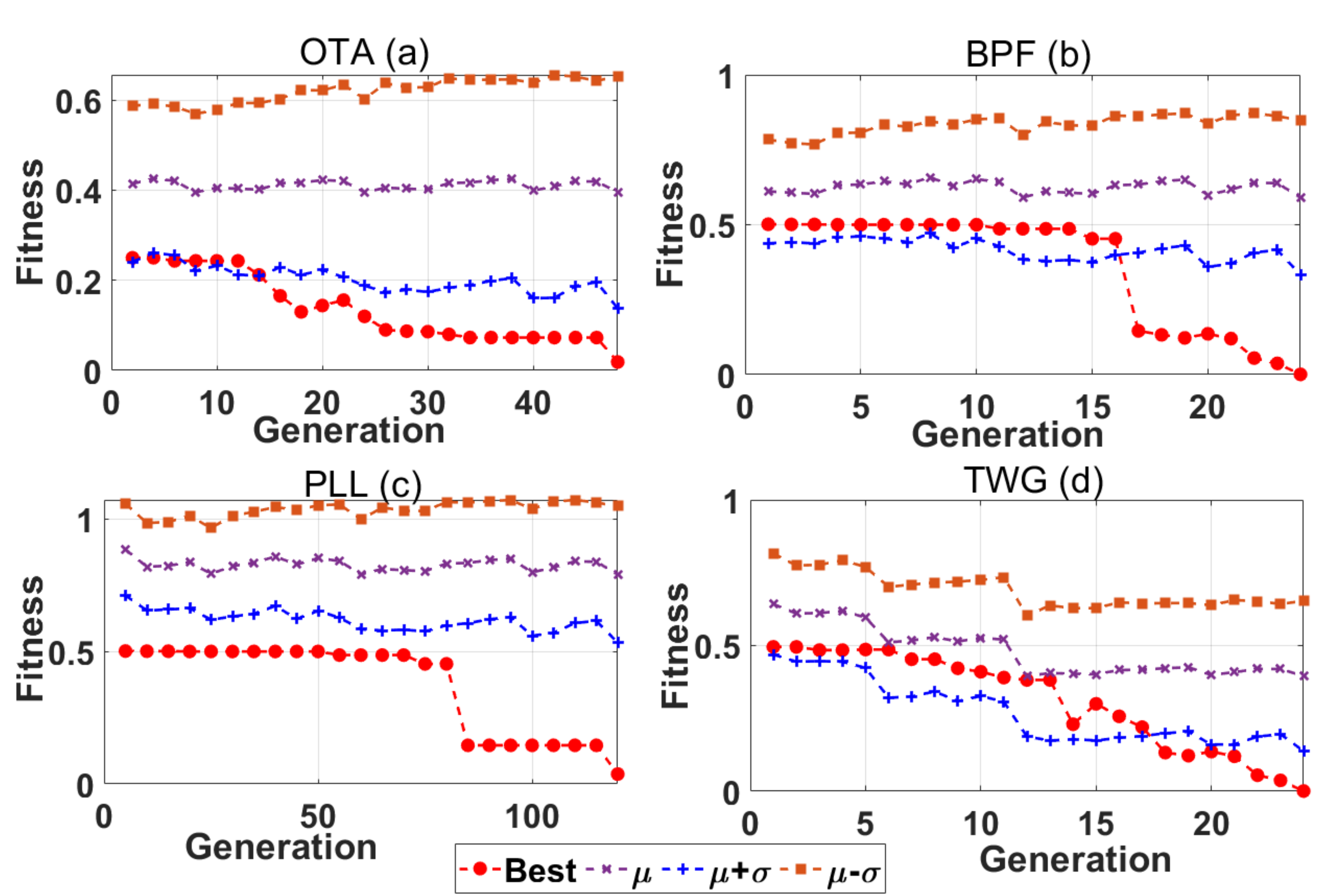}}
\caption{Fitness results across generations of GA for the four different benchmarks (OTA, BPF, PLL, TWG). Note that the attack execution is stopped as soon as the fitness objective is met.}
\label{fig:GA_results_case1}
\end{figure}

Fig.~\ref{fig:GA_results_case1} shows the results of this case over multiple iterations of the GA attack. Note that if the output of the evolved netlist matches the oracle output, then the fitness value is equal to $0$ and GA halts the evolution process. For each generation, the plots show the average fitness value over all chromosomes ($\mu$), and one standard deviation away from the mean values ($\mu\pm \sigma$) as well as the value of the fittest chromosome represented by ``Best".   
Here the best fitness refers to the optimal fitness values in a population at a specific generation, i.e., starts close to $0$ and decreases slowly until GA converges towards the correct solution. 
Below we discuss the results for each benchmark.

\vspace{0.5ex}

\noindent\textbf{OTA.} In an OTA~\cite{benchmark:OTA}, the widths of transistors are obfuscated using additional transistors which are then controlled by the key bits to obfuscate the bias current ($I_{B})$ of the amplifier. For this case, we replace these additional transistors with a single transistor whose width is unknown. The only output parameter the attacker has access to is the frequency response of the amplifier. Using this as our fitness criteria, the GA evolves different widths to calculate the equivalent width ($W_{eff}$) such that the frequency response of this evolved circuit matches that of the unlocked circuit.  
As shown in Fig.~\ref{fig:GA_results_case1}a, the GA reaches the fitness value of $0$ during $48^{th}$ generation of the evolution process. The time taken for GA to extract this value is $33\,s$.  

\begin{figure*}[t]
\subfloat[]{\includegraphics[width=0.2\textwidth,height=0.65\textwidth,keepaspectratio]{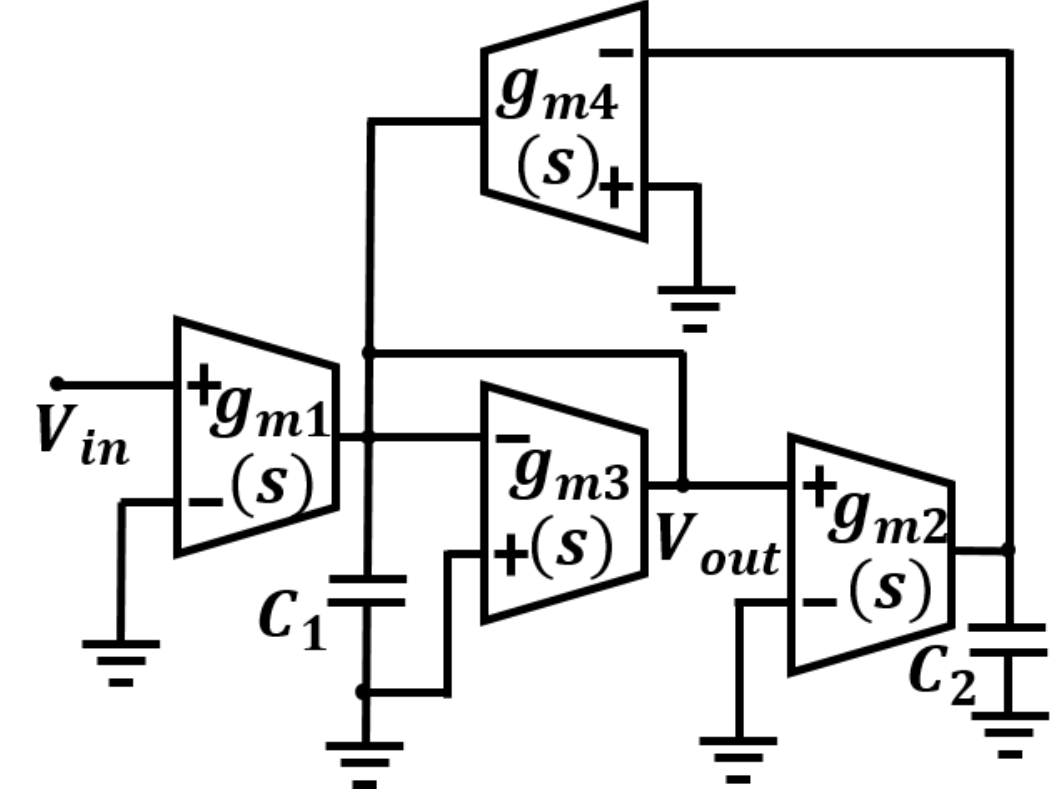}\label{fig_BPF}}
\subfloat[]{\includegraphics[width=0.78\textwidth,height=0.65\textwidth,keepaspectratio]{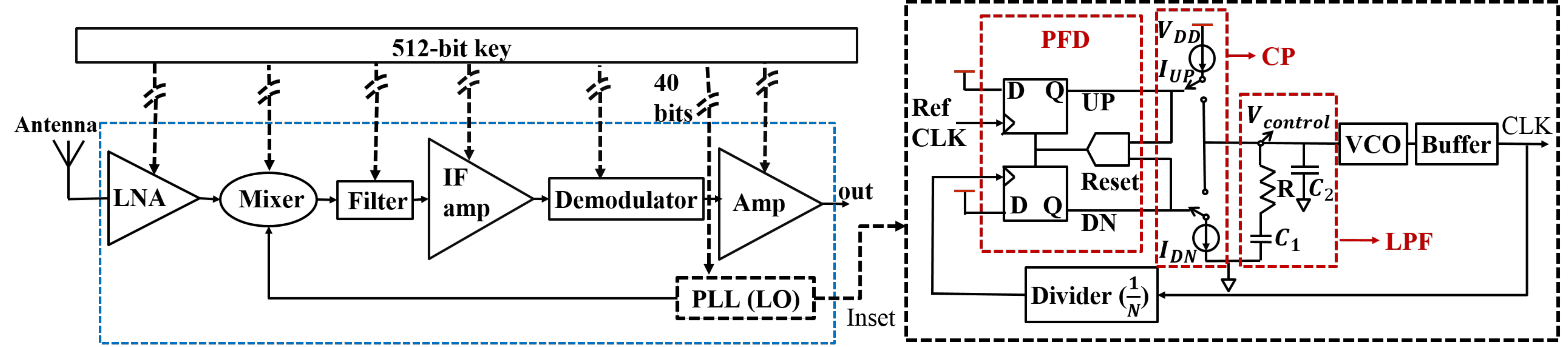}\label{fig_superheterodyne}}
\caption{(a) Second order Gm-C band pass filter (BPF) design; (b) Block diagram of a superheterodyne receiver with an integrated type-II 3rd order PLL (inset).}
\end{figure*}

\vspace{0.5ex}

\noindent\textbf{BPF.} The BPF has two second-order band-pass filters, as shown in Fig.~\ref{fig_BPF}, in cascade~\cite{benchmark:BPF}. In total, the BPF has eight OTAs, out of which six are locked. The key bits are divided across the OTAs. For example, a $33$ bit key is distributed across $6$ OTAs which means six of these current mirrors will be locked by replacing the biasing transistor in each current mirror with multiple transistors~\cite{wang2017thwarting}. This key, like before, obfuscates the bias current ($I_B$) of the amplifier that in turn obfuscates other circuit parameters such as central frequency ($f_c$) and bandwidth ($BW$) of the filter. In this case, the attacker measures the output frequency response of the filter from the oracle. From the obfuscated netlist, the additional transistors are removed and replaced with a single transistor. Then using the output frequency response as our fitness criterion, the $W_{eff}$ is discovered by GA for each of the current mirrors. In this case, the fitness calculation will also be performed according to Equation~\eqref{eq:fitness_simplified}. As shown in Fig.~\ref{fig:GA_results_case1}b, the GA reaches the fitness value of $0$ during $24^{th}$ generation of the evolution process. The best fitness value converges slowly, compared to the case of OTA. This can be because there are six current mirrors whose bias current needs to be appropriate to match the correct output, and there are only a small number of width values that satisfy this condition. At around generation $16$, the GA throws away the lesser-fit parents using the Age-fitness Pareto algorithm as described in Section~\ref{section:experiment_setup}. This results in a sharp improvement in the best fitness value and eventual convergence to the correct $W_{eff}$ by the GA. The entire GA attack completes in $61.7s$.

\vspace{0.5ex}

\noindent\textbf{PLL.} The PLL, as shown in the inset of Fig.~\ref{fig_superheterodyne}, has various interconnected components~\cite{benchmark:PLL}. The locked component of the PLL is an LC-based voltage controlled oscillator (VCO) whose oscillation frequency is a function of the control voltage~\cite{benchmark:VCO}. The locking mechanisms obfuscate the range of this control voltage which in turn obfuscates the operating frequency {$f_{operating}$} of the PLL. The attacker does not know the correct {$f_{operating}$} but can measure the output transient and frequency responses for different inputs from the pins of the unlocked IC.  This means that the number of fitness criterion $m$ will be $2$. Then using Equation~\ref{eq:fitness}, the fitness value will be calculated. 
Fig.~\ref{fig:GA_results_case1}c shows that GA converges towards $W_{eff}$ during $124^{th}$ generation of the evolution process. The graph related to the best fitness values remains steady until the $64^{th}$ generation as GA is trying to converge towards the appropriate locking frequency of the circuit. Similar to the BPF scenario, the Age-fitness Pareto algorithm triggers during the $72^{nd}$ generation to remove lesser fit chromosomes from the population. Then this new generation of individuals helps GA to quickly converge towards the correct solution. The entire GA attack takes $118.3s$ to find $W_{eff}$.

\vspace{0.5ex}

\noindent\textbf{TWG.} TWG consists of two current mirrors that help generate a triangular carrier wave~\cite{benchmark:TWG}. These current mirrors are obfuscated using the combinational locking techniques discussed above, thus obfuscating the frequency ($f_{TRI}$) and the amplitude of the TWG. Similar to before, we first replace the additional transistors in the obfuscated current mirror by a single transistor. Then using the output frequency response as our fitness criterion, we calculate the width value of this single transistor for each current mirror used in the TWG. The fitness value is calculated according to Equation~\eqref{eq:fitness_simplified}. The results as shown in Fig.~\ref{fig:GA_results_case1}d show that during every generation the fitness keeps improving. The average fitness value is also quite low, and the GA converges towards the correct solution during $24^{th}$ generation. In this case, the Age-fitness Pareto algorithm based selection did not trigger. GA extracts the width values in $43.5s$.

\begin{figure}[btp]
\centerline{\includegraphics[width=0.45\textwidth,height=0.45\textwidth,keepaspectratio]{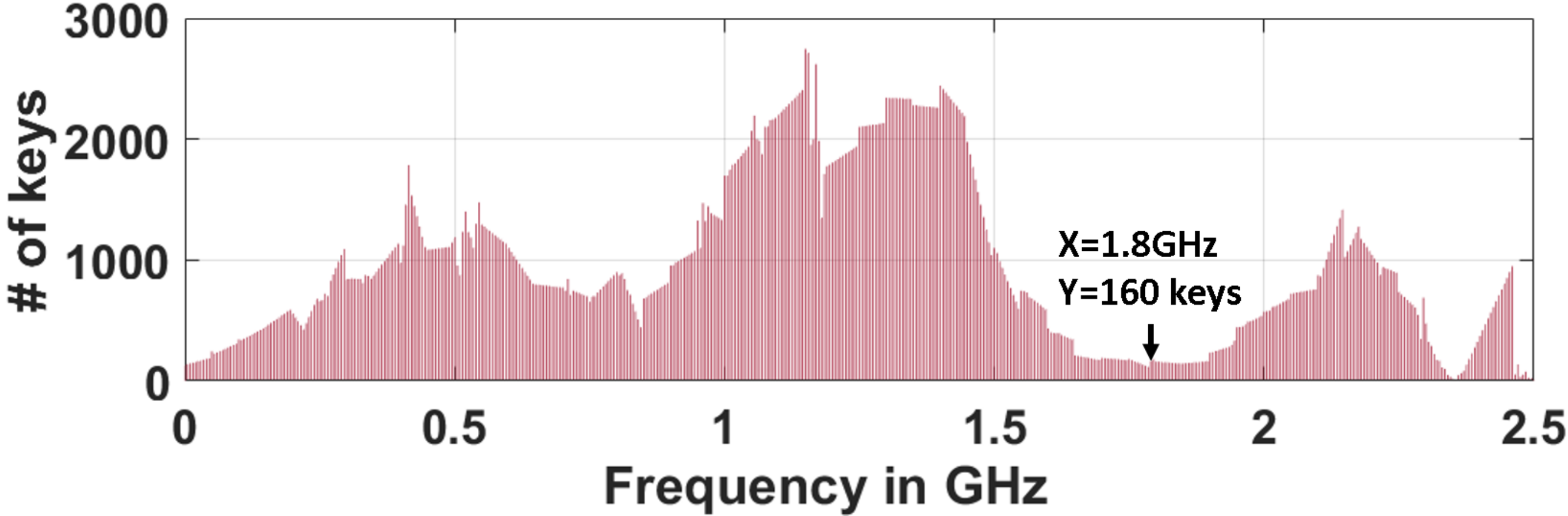}}
\caption{Multiple keys matching oracle output response (locking frequency of PLL=1.8GHz) in PB-Lock technique.}
\label{fig:keyresponse_PLL}
\end{figure}

\begin{table}[t]
\begin{threeparttable}
\caption{The results of GA attack and SMT-based attack on analog circuits locked using SMT-based locking (SMT-lock)~\cite{wang2017thwarting} and parameter-biasing lock (PB-Lock)~\cite{rao2017parameter}. For each circuit benchmark, we show the key length ($k$), the number of keys returned ($K'$), and the time taken for the attack in seconds ($t$).\\
}
\scriptsize
\begin{tabular}{c|c|c|c|c|c||c|c|c|c|}
\cline{2-10}
 &  & \multicolumn{4}{c||}{\cellcolor[HTML]{DAE8FC}\textbf{GA attack}} & \multicolumn{4}{c|}{\cellcolor[HTML]{9698ED}\textbf{SMT attack}} \\ \cline{3-10} 
 &  & \multicolumn{2}{c|}{\textbf{SMT-lock}} & \multicolumn{2}{c||}{\textbf{PB-Lock}} & \multicolumn{2}{c|}{\textbf{SMT-lock}} & \multicolumn{2}{c|}{\textbf{PB-Lock}} \\ \cline{3-10} 
\multirow{-3}{*}{\textbf{\begin{tabular}[c]{@{}c@{}}\end{tabular}}} & \multirow{-3}{*}{\textbf{\begin{tabular}[c]{@{}c@{}}k\end{tabular}}} & \textbf{\begin{tabular}[c]{@{}c@{}}$K'$\end{tabular}} & \textbf{\begin{tabular}[c]{@{}c@{}}$t$ [s]\end{tabular}} & \textbf{\begin{tabular}[c]{@{}c@{}}$K'$\end{tabular}} & \textbf{\begin{tabular}[c]{@{}c@{}}$t$ [s]\end{tabular}} & \textbf{\begin{tabular}[c]{@{}c@{}}$K'$\end{tabular}} & \textbf{\begin{tabular}[c]{@{}c@{}}$t$ [s]\end{tabular}} & \textbf{\begin{tabular}[c]{@{}c@{}}$K'$\end{tabular}} & \textbf{\begin{tabular}[c]{@{}c@{}}$t$ [s]\end{tabular}} \\ \hline
 \multicolumn{1}{|c|}{} & 16 & 1 & 17.1 & 1\tnote{*} & 11.2 & 1 & 1.3 & 7 & 1.23 \\ \cline{2-10} 
\multicolumn{1}{|c|}{} & 32 & 1 & 22.6 & 1\tnote{*} & 19.3 & 3 & 1.7 & 10 & 1.71 \\ \cline{2-10} 
\multicolumn{1}{|c|}{} & 40 & 1 & 27 & 1\tnote{*} & 21.4 & 7 & 2.3 & 17 & 2.2 \\ \cline{2-10} 
\multicolumn{1}{|c|}{\multirow{-4}{*}{OTA}} & 64 & 1 & 43.8 & 1\tnote{*} & 38.3 & 8 & 4.1 & 22 & 4 \\ \hline
\multicolumn{1}{|c|}{} & 16 & 1 & 31.8 & 1\tnote{*} & 29 & 4 & 7.1 & 12 & 3.4 \\ \cline{2-10} 
\multicolumn{1}{|c|}{} & 32 & 1 & 63.4 & 1\tnote{*} & 47.5 & 5 & 9.3 & 19 & 4.7 \\ \cline{2-10} 
\multicolumn{1}{|c|}{} & 40 & 1 & 99.6 & 1\tnote{*} & 41 & 7 & 14.7 & 21 & 11 \\ \cline{2-10} 
\multicolumn{1}{|c|}{\multirow{-4}{*}{BPF}} & 64 & 1 & 134.3 & 1\tnote{*} & 111.4 & 11 & 17.7 & 23 & 13.6 \\ \hline
\multicolumn{1}{|c|}{} & 16 & 1 & 36 & 1 & 33.4 & 3 & 13.5 & 7 & 13 \\ \cline{2-10} 
\multicolumn{1}{|c|}{} & 32 & 1 & 71.5 & 1 & 64.8 & 8 & 21 & 16 & 20.7 \\ \cline{2-10} 
\multicolumn{1}{|c|}{} & 40 & 1 & 107.2 & 1 & 105.4 & 13 & 37.7 & 22 & 39 \\ \cline{2-10} 
\multicolumn{1}{|c|}{\multirow{-4}{*}{PLL}} & 64 & 1 & 144.5 & 1\tnote{*} & 121.6 & 17 & 41.3 & 24 & 40.4 \\ \hline
\multicolumn{1}{|c|}{} & 16 & 1 & 22.7 & 1\tnote{*} & 18.3 & 1 & 2.3 & 4 & 1.8 \\ \cline{2-10} 
\multicolumn{1}{|c|}{} & 32 & 1 & 54.2 & 1\tnote{*} & 32.9 & 1 & 3.6 & 9 & 2.7 \\ \cline{2-10} 
\multicolumn{1}{|c|}{} & 40 & 1 & 89.4 & 1\tnote{*} & 63.4 & 6 & 4.4 & 11 & 3.3 \\ \cline{2-10} 
\multicolumn{1}{|c|}{\multirow{-4}{*}{TWG}} & 64 & 1 & 117.6 & 1\tnote{*} & 110.3 & 8 & 8.1 & 22 & 7.4 \\ \hline
\multicolumn{1}{|c|}{\textbf{Avg.}} & \textbf{9.5} & \textbf{1} & \textbf{67.6} & \textbf{1} & \textbf{54.3} & \textbf{6.4} & \textbf{11.8} & \textbf{15.3} & \textbf{10.6} \label{tab:attack_time}\\ \hline
\end{tabular}
\end{threeparttable}
\begin{tablenotes}
\item[*] \footnotesize{ $*$ GA halts the evolution process after returning a single key. The output for this key matches the oracle output. However, this might not be the actual key which the circuit benchmark was locked with.}
\end{tablenotes} 
\end{table}
\normalsize

\vspace{0.5ex}

\noindent\textbf{Summary of Case 1.} We successfully used GA to find the value of the obfuscated parameter that matches the oracle output for all four different circuit benchmarks. This takes $2$ minutes at worst. This calculated value can also be used as an additional fitness criterion to reduce the search space of extracting a correct key from $2^{k}$ to linear, which we will demonstrate on the PLL and the superheterodyne receiver in Sections~\ref{sec_case2_GA} and~\ref{sec_Super_R}.

\subsection{Results of Case 2}
\subsubsection{GA Attack Results} \label{sec_case2_GA}
Before discussing the results, it is worth revisiting the major difference between SMT-Lock and PB-Lock. In the former, the key is made to be unique while in the latter it is selected randomly. Thus, in the case of PB-Lock, the GA might return a different key than the one randomly selected by the locking scheme. Nevertheless, the key returned by GA results in a circuit that matches the oracle. This illustrates a significant security flaw in PB-Lock and the advantage of the GA attack. Fig.~\ref{fig:keyresponse_PLL} shows the number of keys that match the oracle output after simulating over $500$ thousand keys for the PLL locked with a $40$-bit key. Even for a highly parameter-sensitive circuit like PLL, there are over $160$ keys that return the exact locking frequency $f_{locking}$ of $1.8 GHz$. In order to speed up the attacks in Case 2 for PLL, we use the \textit{two pass variant} (described in Section~\ref{sec:2pass}) with multiple fitness criteria including the value of $W_{eff}$ obtained from Case 1. Multiple-fitness criteria also helps in tackling the problem of non-monotonicity~\cite{zebulum2018evolutionary}. While attacking the circuit locked with SMT-based locking technique~\cite{wang2017thwarting}, GA will take slightly longer to converge towards the correct key. The time taken to extract the correct key mainly depends on the complexity of the circuit, key length $k$, and number of fitness criterion used. 

\begin{figure*}[t]
\subfloat[]{\includegraphics[width=0.48\textwidth,height=0.48\textwidth,keepaspectratio]{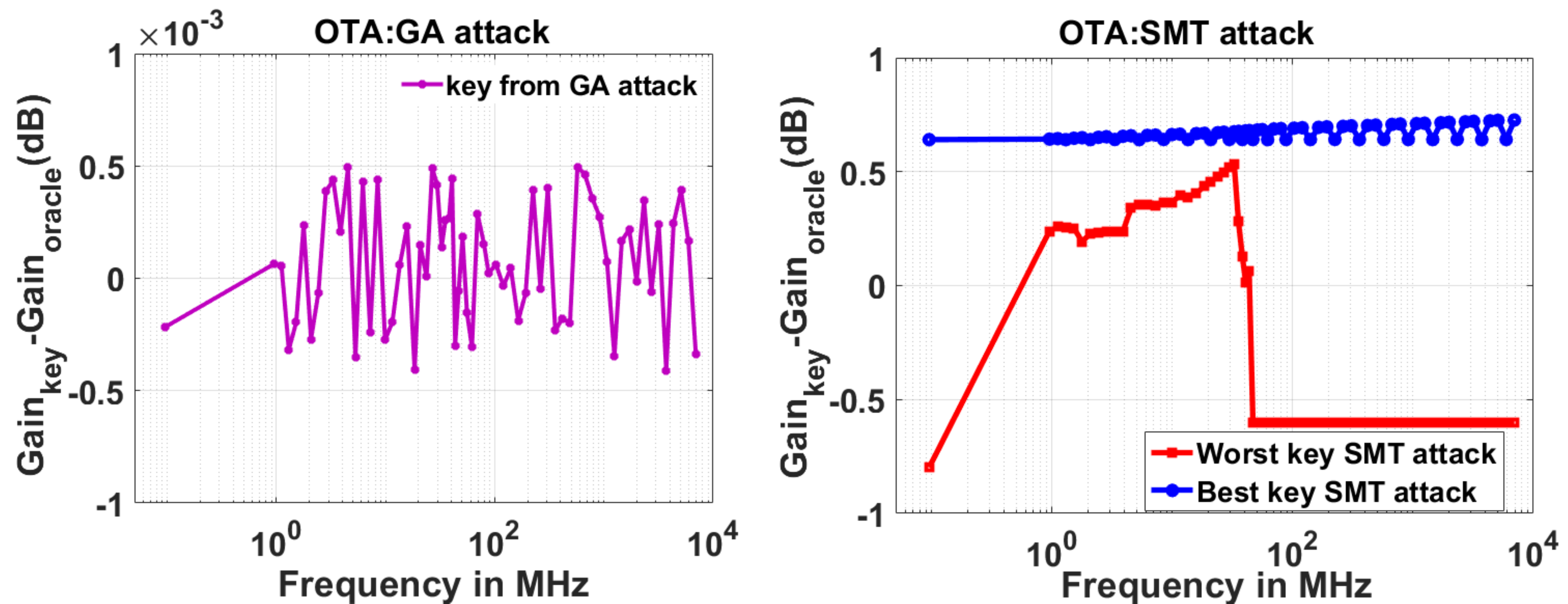}\label{fig:OTA_GA_SMT_comp}}
\subfloat[]{\includegraphics[width=0.43\textwidth,height=0.43\textwidth,keepaspectratio]{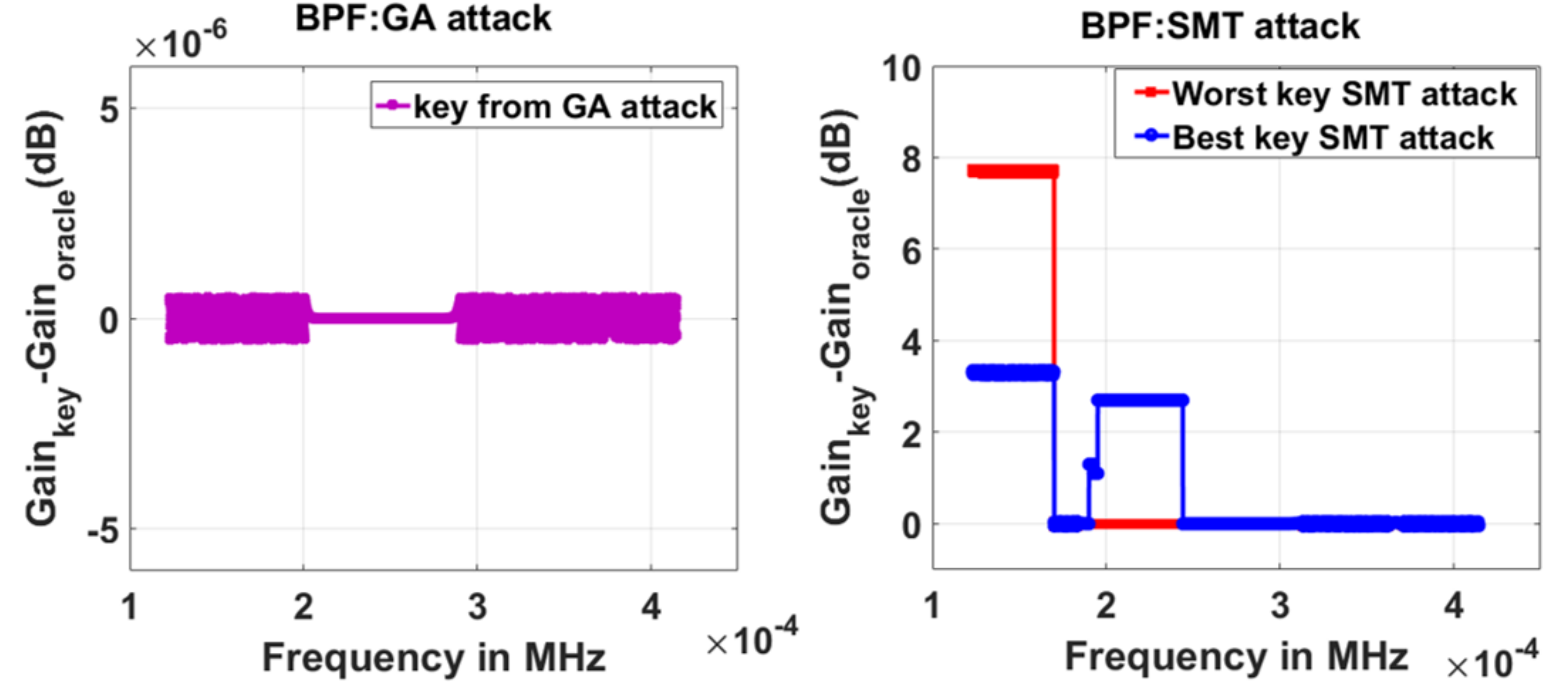}\label{fig:BPF_GA_SMT_comp}}\\
\subfloat[]{\includegraphics[width=0.43\textwidth,height=0.43\textwidth,keepaspectratio]{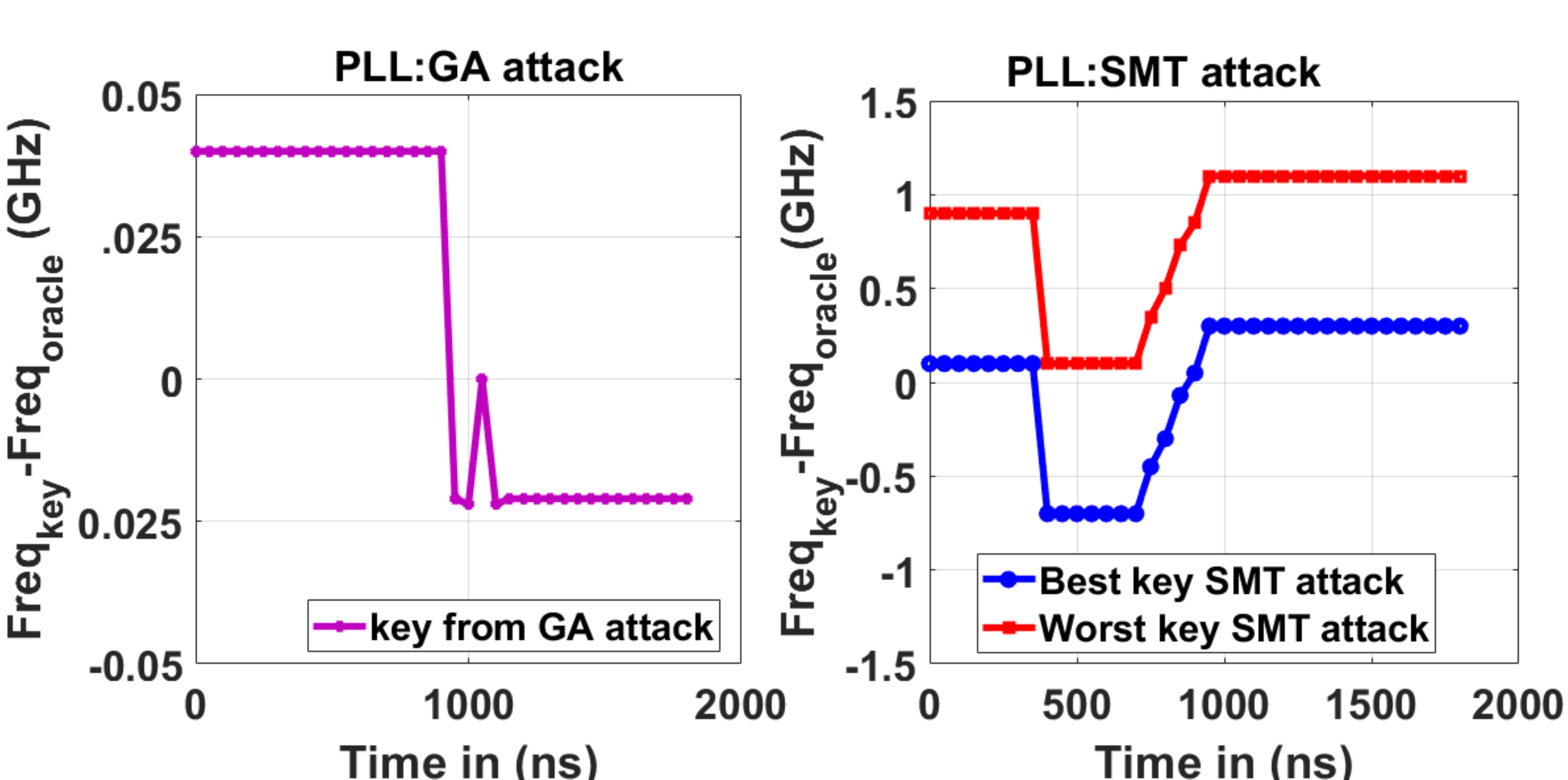}\label{fig:PLL_GA_SMT_comp}}
\subfloat[]{\includegraphics[width=0.48\textwidth,height=0.48\textwidth,keepaspectratio]{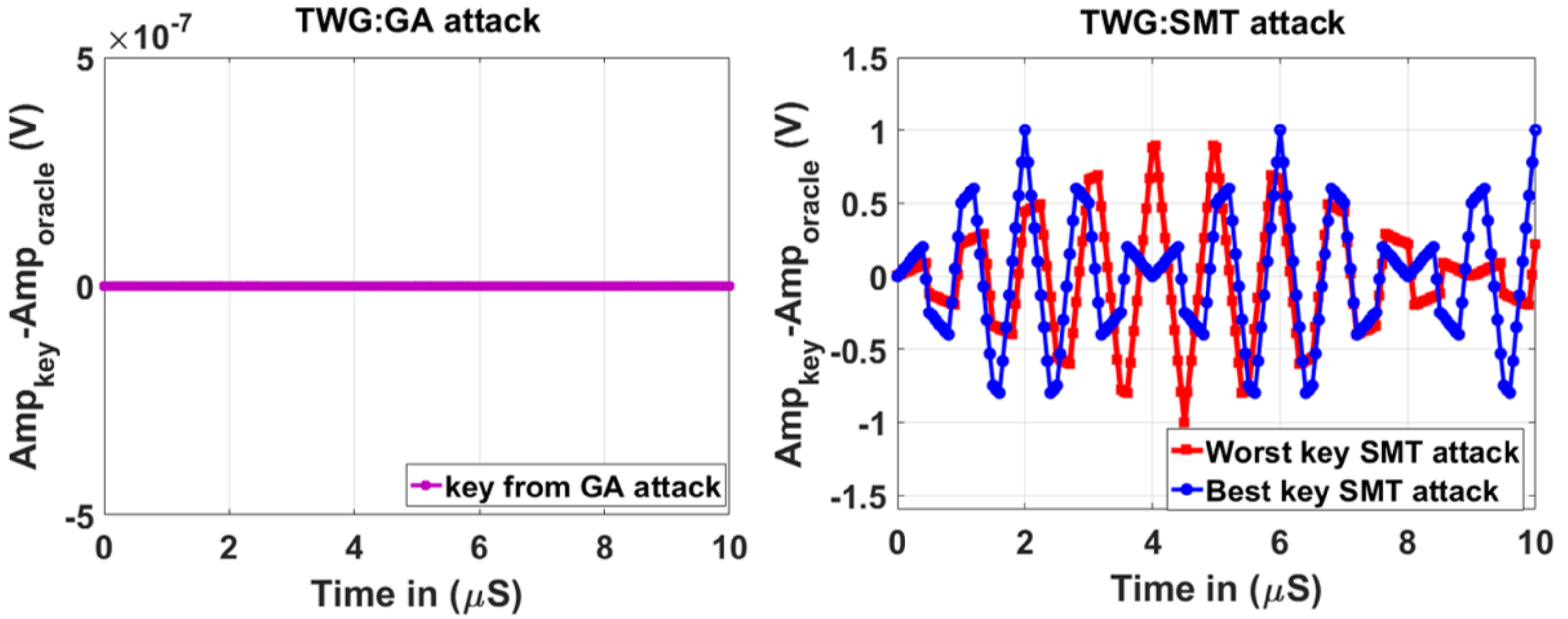}\label{fig:TWG_GA_SMT_comp}}
\caption{Difference between oracle and unlocked chip with keys recovered through GA and SMT attacks for the benchmarks (a) OTA, (b) BPF, (c) PLL, (d) TWG. GA attack provides one key while the SMT attack provides multiple keys. The best and worst SMT key results are shown for simplicity.}
\label{fig:results_Case2}
\end{figure*}

The number of keys extracted  ($K'$) and the time taken ($t$) by GA attack for the four different circuit benchmarks locked by SMT-Lock and PB-Lock are summarized in the lefthand side of Table~\ref{tab:attack_time}. In the case of SMT-lock, GA always returns the correct key. This is because there is only one key which sets the current bias properly such that the output response is unique for that key. By using multiple fitness criteria (including the value of the equivalent $W_{eff}$ obtained from Case 1), the search space is reduced from exponential ($2^{k}$) to linear. For PB-Lock, GA attack returns a single key which does not necessarily match the locking key, but still unlocks the circuit as described above.
For both SMT-Lock and PB-Lock, the GA attack completes in less than 2.5 minutes for all benchmarks and $k$. Further, it is worth noting that the attack time increases linearly with respect to $k$.

\subsubsection{SMT Attack Results}
As described in Section~\ref{subsection:SMT_attack}, the SMT-attack requires that the adversary know the desired current value of the current mirrors and have good analog design experience to formulate the right equations. 

The righthand side of Table~\ref{tab:attack_time} shows the results for SMT-based attack on circuits locked with SMT-Lock and PB-Lock. SMT-based attack takes less amount of time than GA but never returns a single key (even in the case of attacking SMT-lock). Thus, a set of output responses need to be collected by applying each key to the locked netlist and compared to the output of the oracle. In other words, the SMT attack does involve a certain amount of brute force to be performed which depending on the complexity of the circuit and the skill of the attacker can take exponential time. 
More precisely, suppose that a set of $K'$ keys are delivered by the SMT solver. 
If all possible keys are checked by using brute force, the time complexity of this inevitable post-processing step is $O(K'2^{k})$. 

\subsubsection{Performance of Keys from GA vs. SMT Attack}
Fig.~\ref{fig:results_Case2} shows the difference in the output performance of the circuit locked with a $64$-bit key using parameter-biasing technique~\cite{rao2017parameter} after applying keys extracted from the GA and SMT attacks. As discussed earlier, GA returns just a single key while SMT returns a set of keys. Below we will explain the results and compare the performance for each benchmark with key returned from GA attack and with best and worst keys returned from SMT attack.

\vspace{0.5ex}

\noindent\textbf{OTA} is characterized by gain $G=41dB$ and unity gain BW = $1.2GHz$. As shown in Fig.~\ref{fig:OTA_GA_SMT_comp}, the key extracted from the GA attack shows a very small difference in the order of $10^{-3}$dB while both keys returned from the SMT attack show a more significant difference in the performance value. The OTA frequency response using the best key from SMT attack shows a difference in gain of about $0.6dB$ throughout the frequency range compared to the oracle output. While the frequency response with the worst key shows a higher difference in gain compared to the oracle.

\vspace{0.5ex}

\noindent\textbf{BPF} is characterized by $G=0dB$, central frequency $f_c= 250kHz$, and $BW = 150kHz$. As shown in Fig.~\ref{fig:BPF_GA_SMT_comp}, both the keys returned from the SMT attack show significant difference in gain and BW while the key returned by GA attack shows little to no difference.

\vspace{0.5ex}

\noindent\textbf{PLL} is characterized by locking frequency $f_{locking}=1.8GHz$ and settling time $T_s = 920ns$. As shown in Fig.~\ref{fig:PLL_GA_SMT_comp}, the circuit using the key returned from the GA attack has $f_{locking}=1.778GHz$ and settling time $T_s=923ns$, which closely matches the oracle output. PLL using the best key returned from the SMT attack has $f_{locking}=1.83GHz$ and settling time $T_s = 953ns$. PLL using one of the worst keys returned from the SMT attack has $f_{locking} =2.92GHz$ and $T_s = 930ns$.

\vspace{0.5ex}

\noindent\textbf{TWG} is characterized by amplitude of $1V$ and time period of $2{\mu}s$. As shown in Fig.~\ref{fig:TWG_GA_SMT_comp}, locked TWG with the key returned from GA attack shows no difference in the output response. However, the keys returned from SMT attack show difference in time period and amplitude. Although the TWG with the keys returned from the SMT-attack has amplitude of $1V$, they have different time-periods compared to the unlocked TWG which then results in a difference in transient response compared to the oracle. TWG with the best key returned from the SMT-attack has a time period of $1.85{\mu}s$ while TWG with the worst key has a time period of $2.28{\mu}s$.

\subsubsection{GA Attack on Locked Superheterodyne Receiver} \label{sec_Super_R}

A superheterodyne receiver is one of the most commonly used RF circuits for modern-day radio receivers. To protect its IP, PB-Lock uses a $512$-bit key to obfuscate the performance of the receiver with $40$-bit key to lock the PLL, see ~\cite{rao2017parameter}. Considering the key length and the number of interconnected components as shown in Fig.~\ref{fig_superheterodyne}, it should take exponential time to extract the correct key. Further. in previous experiments, the attacker had direct access to the output of the locked element of the circuit. Here, however, the only accessible output of the oracle is the receiver output (i.e., $out$ of Amp in Fig.~\ref{fig_superheterodyne}). 

Here, we attempt to unlock the the PLL using the two-pass variants. First, we employ GA to extract the equivalent value of the PLL's obfuscated parameter (i.e., $W_{eff}$). While this is  similar to Case 1, the output of the receiver is used for the fitness criteria instead of the PLL output. Then, we use $W_{eff}$ as well as the frequency response of the receiver as our fitness criteria to extract the correct key of the locked PLL, i.e., the number of fitness criterion $m=2$ and Equation~\eqref{eq:fitness} becomes
\begin{equation}
\small
F=\sum_{i=1}^{n}(E_{i,1} - O_{i,1})^2 + \sum_{i=1}^{n}(E_{i,2} - O_{i,2})^2,\label{eq:lock_receiver}
\end{equation}
where $E_{i,1}$ and $E_{i,2}$ refers to the target value of the equivalent $W_{eff}$ and the frequency response from the oracle respectively while $O_{i,1}$ and $O_{i,2}$ refers to the width value obtained for a specific key and the frequency response obtained after simulating the netlist of the receiver for a specific key.

\begin{figure}[tpb]
\centerline{\includegraphics[width=0.5\textwidth,height=0.5\textwidth,keepaspectratio]{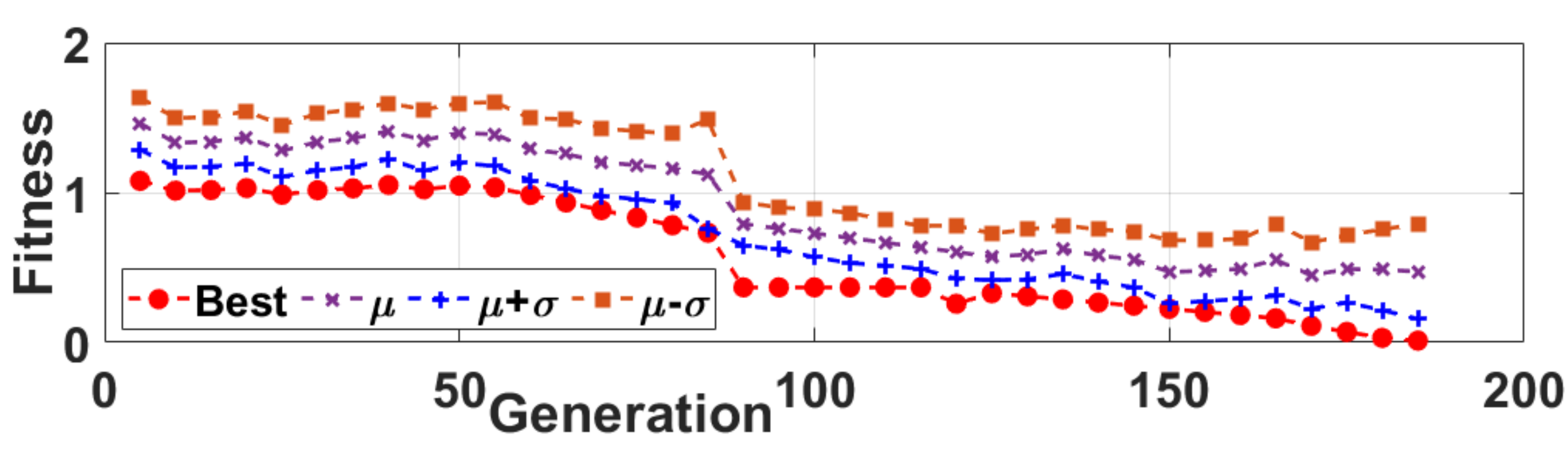}}
\caption{Fitness results across generations for superheterodyne receiver. 
}
\label{fig:superheterodyne_fitness}
\end{figure}

The key locking the PLL is extracted by GA in $314.3$ seconds in the $181^{st}$ generation as can be seen in Fig~\ref{fig:superheterodyne_fitness}. The key returned from GA matches the one used to lock the PLL. This shows that the correct key can be extracted \textit{even without using the immediate output of the PLL}. The rest of the keys locking the receiver can be extracted more easily by GA (not shown for brevity) as the receiver's performance is most heavily dependent on the PLL. 

\vspace{0.5ex}

\noindent\textbf{Summary of Case 2.} Salient points from this section are: 
\begin{itemize}
    \item GA successfully extracted the key from the benchmarks locked with different key sizes of both SMT-lock~\cite{wang2017thwarting} and PB-lock~\cite{rao2017parameter}. GA returned a single key whose output matched exactly with the oracle output. This was accomplished at worst under $2.5$ minutes. 
    \item For comparison purposes, the same benchmarks were attacked using SMT-solver which returned multiple keys. The SMT-attack returned the set of keys faster, at worst under a minute, but the time complexity of the inevitable brute force to obtain the correct key is $O(K^{'}2^{L})$.
    \item The GA was able to unlock the PLL of a superheterodyne receiver using the output of the receiver instead of the PLL's immediate output. The key was successfully extracted in about $5$ minutes.
\end{itemize}

\subsection{Discussion and Final Takeaways} \label{sec_takeway}
In this section, we have described the experimental setup and the results obtained from the proposed GA based attack and compared the results with SMT-based attack on SMT-Lock and PB-Lock. Below we list the advantages of our attack and its extension to other known analog locking techniques. 

\vspace{0.5ex}

\noindent\textbf{Advantages of GA vs. SMT Attack.}
\begin{itemize}
    \item GA attack is a straightforward approach that does not require any prior knowledge of the locking technique or the circuit benchmark. With just the oracle and the obfuscated netlist, the attacker can extract the correct key from the locked circuit as shown in Case 2 experiments.
    \item Rather than find the correct key, the attacker can also find the value of the obfuscated parameter which takes linear time as shown in Case 1 experiments. This value can further be used to reduce the search space for finding the correct key from $2^k$ to linear time as shown in Case 2 experiments.
    \item SMT-solver, on the other hand, uses theoretical circuit-equations, equations describing the locking technique, and the oracle to attack the circuit and return a set of keys whose performances did not match the oracle output as closely as the GA attack, see Fig.~\ref{fig:results_Case2} and Table~\ref{tab:attack_time}.
\end{itemize}

\vspace{0.5ex}

\noindent\textbf{Applying GA Attack to Other Analog Locking Schemes.} It is important to note that the GA attack is generalizable, and should be capable of defeating schemes other than SMT-Lock and PB-Lock. We make the following inferences. 
\begin{itemize}
    \item The Case 1 experiments demonstrate that GA attack can discover obfuscated parameters, such as $W$. Thus, it should also be able to obtain the value of the obfuscated threshold voltages $V_{th}$ used in~\cite{multi_threshold_locking} by employing multiple fitness criterion for different output parameters (such as frequency response and transient response).
    \item The Case 2 experiments (most notably, superheterodyne receiver) demonstrate that the GA attack can extract internal keys and parameters using just a netlist and the oracle. We believe that 
    it can, therefore, be used to break the neural network-based analog circuit locking described in~\cite{volanis2019analog} by extracting either the neural network's analog key or the internal biases controlled by that.
\end{itemize}

\section{Conclusion and Future Work}
In this paper, we showed how an evolutionary approach like the genetic algorithm (GA) can easily break the security of the analog locking techniques. We successfully extracted the value of the obfuscated parameter as well as the correct key from the circuits locked with popular techniques in the literature. 
Given the oracle output and the obfuscated netlist, we also showed how the attacker can easily estimate the value of the obfuscated parameter to reduce the search space of finding the correct key. In future work, we shall demonstrate the GA attack on other locking techniques. 

\small
\bibliographystyle{ieeetr}
\balance
\bibliography{conference_041818}
\end{document}